Research Article

# Exoplanet Biosignatures: A Framework for Their Assessment

David C. Catling,[1,2] Joshua Krissansen-Totton,[1,2] Nancy Y. Kiang,[3] David Crisp,[4] Tyler D. Robinson,[5] Shiladitya DasSarma,[6] Andrew J. Rushby,[7] Anthony Del Genio,[3] William Bains,[8] and Shawn Domagal-Goldman[9]

**Abstract**

Finding life on exoplanets from telescopic observations is an ultimate goal of exoplanet science. Life produces gases and other substances, such as pigments, which can have distinct spectral or photometric signatures. Whether or not life is found with future data must be expressed with probabilities, requiring a framework of biosignature assessment. We present a framework in which we advocate using biogeochemical "Exo-Earth System" models to simulate potential biosignatures in spectra or photometry. Given actual observations, simulations are used to find the Bayesian likelihoods of those data occurring for scenarios with and without life. The latter includes "false positives" wherein abiotic sources mimic biosignatures. Prior knowledge of factors influencing planetary inhabitation, including previous observations, is combined with the likelihoods to give the Bayesian posterior probability of life existing on a given exoplanet. Four components of observation and analysis are necessary. (1) Characterization of stellar (*e.g.*, age and spectrum) and exoplanetary system properties, including "external" exoplanet parameters (*e.g.*, mass and radius), to determine an exoplanet's suitability for life. (2) Characterization of "internal" exoplanet parameters (*e.g.*, climate) to evaluate habitability. (3) Assessment of potential biosignatures within the environmental context (components 1–2), including corroborating evidence. (4) Exclusion of false positives. We propose that resulting posterior Bayesian probabilities of life's existence map to five confidence levels, ranging from "very likely" (90–100%) to "very unlikely" (<10%) inhabited. Key Words: Bayesian statistics—Biosignatures—Drake equation—Exoplanets—Habitability—Planetary science. Astrobiology 18, xxx–xxx.

## 1. Introduction

I<small>N THE FUTURE</small>, if unusual combinations of gases or spectral features are detected on a potentially habitable exoplanet, consideration will undoubtedly be given to the possibility of their biogenic origin. But the extraordinary claim of life should be the hypothesis of last resort only after all conceivable abiotic alternatives are exhausted. The possibility of false positives—when a planet without life produces a spectral or photometric feature that mimics a biosignature—is a lesson learned from consideration of oxygen and ozone ($O_3$) as biosignatures (Meadows *et al.*, 2018, in this issue). If life is suggested by remote data, the discovery will always have some uncertainty and so the extent to which data suggest the presence of life should be assigned a probability. Following this approach, we outline a general framework for detecting and verifying biosignatures in exoplanet observations. Additional research required to implement biosignature assessment is discussed elsewhere (Walker *et al.*, 2018, in this issue), as are missions and observatories that will eventually acquire the data (Fujii *et al.*, 2018, in this issue).

In its broadest definition, a *biosignature* is any substance, group of substances, or phenomenon that provides evidence of life (reviewed by Schwieterman *et al.*, 2018, in this issue). The objective of the framework we describe is to identify the information and general procedures required to quantify and increase the confidence that a suspected biosignature detected on an exoplanet is truly a detection of life.

[1]Astrobiology Program, Department of Earth and Space Sciences, University of Washington, Seattle, Washington.
[2]Virtual Planetary Laboratory, University of Washington, Seattle, Washington.
[3]NASA Goddard Institute for Space Studies, New York, New York.
[4]MS 233-200, Jet Propulsion Laboratory, California Institute of Technology, Pasadena, California.
[5]Department of Astronomy and Astrophysics, University of California, Santa Cruz, California.
[6]Department of Microbiology and Immunology, School of Medicine, and Institute of Marine and Environmental Technology, University of Maryland, Baltimore, Maryland.
[7]NASA Ames Research Center, Moffett Field, California.
[8]Department of Earth, Atmospheric and Planetary Science, Cambridge, Massachusetts.
[9]NASA Goddard Space Flight Center, Greenbelt, Maryland.

Here, we restrict our framework to the type of biosignatures that might be detected in emitted, transmitted, or reflected light from exoplanets as a result of the biogeochemical products of a biosphere. We do not consider technosignatures such as search for extraterrestrial intelligence (SETI) radio or visible broadcasts from technological civilizations, infrared (IR) excess from Dyson spheres, or other so-called megastructures, and so on. Such specialist matters are reviewed elsewhere (*e.g.*, Tarter, 2007; Wright, 2017).

## 2. A Bayesian Framework for Biosignature Assessment and Life Detection

A Bayesian approach is an appropriate technique to provide a best-informed probability about a hypothesis when dealing with incomplete information. In assessing biosignatures, we face the problem of assigning a probability to whether an exoplanet is inhabited based on remote observations that will always be limited, especially given that *in situ* observations of even the nearest exoplanets are very far in the future (*e.g.*, Lubin, 2016; Heller and Hippke, 2017; Manchester and Loeb, 2017). As we shall see, Bayes' theorem quickly reveals our current ignorance about life on exoplanets. However, exposure of what we do not know points to what research needs to be done and which observations would increase confidence in possible life detection.

Our goal is to calculate the probability that the hypothesis of life existing on an exoplanet is true given observational data that show possible biosignatures. In terms of Bayesian statistics [*e.g.*, see Stone (2013) and Sivia and Skilling (2006) for introductions], we seek the conditional probability $P(\text{life} \,|\, \text{data}, \text{context})$ where "life | data, context" means "the hypothesis of life existing on an exoplanet *given* the observed data that may contain biosignatures *and* the exoplanet's context." In this expression, "|" means "given" and the comma means "and." The "data" are spectra and/or photometry from an exoplanet with possible biosignatures. The "context" consists of the stellar and planetary parameters that are relevant to the possibility of life producing the spectral and/or photometric data, as shown in Figure 1. We take the presence of life to mean the existence of a surface biosphere that is remotely detectable, which we treat as a binary variable, that is, either an exoplanet biosphere is present or it is not. Of course, this approach cannot take into account hidden subsurface biospheres [*e.g.*, as postulated for Europa's ocean (*e.g.*, Chyba and Phillips, 2001)] or cryptic surface biospheres that are too meager to have any effect on an exoplanet surface or atmosphere that is detectable with remote sensing. Hidden or cryptic biospheres are not practical candidates for life detection on exoplanets, so their consideration is not relevant to the empirical perspective of this article.

With the aforementioned assumptions, we now develop an expression that forms the basis of a statistical framework for assessing the probability of life on an exoplanet. We start by assuming relatively little familiarity with Bayesian statistics because this article is intended to be accessible to a wide audience across disparate disciplines, which is needed for astrobiology.

Bayes' theorem, in its standard form, is

$$P(\text{hypothesis} \,|\, \text{evidence}) = \underbrace{\frac{P(\text{evidence} \,|\, \text{hypothesis})}{P(\text{evidence})}}_{\text{support the evidence provides for the hypothesis}} \times \underbrace{P(\text{hypothesis})}_{\text{the prior}}. \quad (1)$$

Here, one reads "hypothesis" as the "hypothesis being true" and the $P$'s are probabilities. The theorem gives $P(\text{hypothesis} \,|\, \text{evidence})$, the probability of the hypothesis being true given the evidence, as a function of the likelihood of the evidence occurring if the hypothesis is true, $P(\text{evidence} \,|\, \text{hypothesis})$ divided by a so-called marginal likelihood, $P(\text{evidence})$. The ratio on the right-hand side is weighted by a prior probability for the hypothesis being true, $P(\text{hypothesis})$.

In the case of a binary hypothesis that is either true or false, and $P(\text{evidence}) > 0$, an extended version of Bayes theorem is

$$P(\text{hypothesis} \,|\, \text{evidence}) = \frac{P(\text{evidence} \,|\, \text{hypothesis})P(\text{hypothesis})}{P(\text{evidence} \,|\, \text{hypothesis})P(\text{hypothesis}) + P(\text{evidence} \,|\, \text{hypothesis false})P(\text{hypothesis false})}. \quad (2)$$

This form of Bayes' theorem is the one used in this article.

We make the extended Bayes' theorem specific to the case of the hypothesis of life existing on an exoplanet. Then, the posterior probability we seek, as mentioned earlier, is $P(\text{life} \,|\, \text{data}, \text{context})$, which is the probability of life on an exoplanet given spectral or photometric "data" of the exoplanet and the "context," which consists of all the known exoplanet system or stellar properties. By Bayes' extended theorem in Eq. 2, and the rules of conditional probability, we have

$$P(\text{life} \,|\, \text{data}, \text{context}) = \frac{P(\text{data}, \text{context} \,|\, \text{life})P(\text{life})}{P(\text{data}, \text{context} \,|\, \text{life})P(\text{life}) + P(\text{data}, \text{context} \,|\, \text{no life})P(\text{no life})}$$

$$= \frac{P(\text{data} \,|\, \text{context}, \text{life})P(\text{context} \,|\, \text{life})P(\text{life})}{P(\text{data} \,|\, \text{context}, \text{life})P(\text{context} \,|\, \text{life})P(\text{life}) + P(\text{data} \,|\, \text{context}, \text{no life})P(\text{context} \,|\, \text{no life})P(\text{no life})}. \quad (3)$$

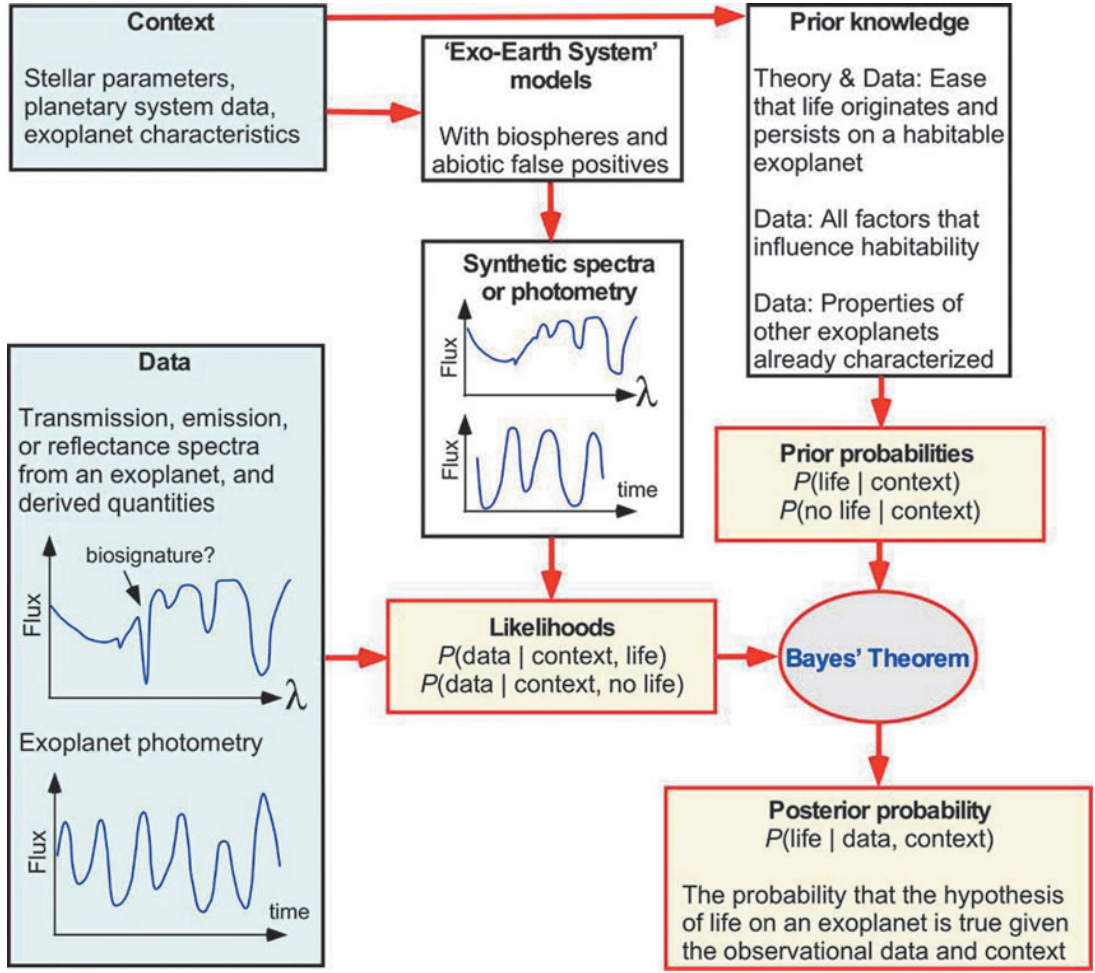

**FIG. 1.** A Bayesian framework for biosignature assessment. Spectral and/or photometric data that may contain biosignatures are used with models to find likelihoods given the context of the exoplanet, for example, its astrophysical environment. One likelihood is the conditional probability of those data occurring given the context and the hypothesis that the exoplanet has life. Another likelihood is the probability of the data occurring given the context and the hypothesis that the exoplanet has no life. These two likelihoods are weighted by prior knowledge to provide a best-informed (posterior) probability that the exoplanet has life given the spectral and/or photometric data and context. Blue boxes signify data acquisition. Yellow boxes contain conditional probabilities and prior probabilities that are part of Bayes' Theorem (gray oval), which is expressed in Eq. 7 in the text.

If a planet has life, then it gives information about context, for example, the planet is not around an early O-type star that only lasts a few million years (Weidner and Vink, 2010) or the planet is not orbiting at a radial distance from a star where temperatures would exceed the stability limit for biomolecules. Thus, from an information point of view, the presence of life even provides information about stellar radiation, for example. To put this another way,

$$P(\text{context} \mid \text{life}) \neq P(\text{context} \mid \text{no life}). \quad (4)$$

Using the laws of conditional probability, we can derive the following expressions:

$$P(\text{context} \mid \text{life}) = \frac{P(\text{life} \mid \text{context})P(\text{context})}{P(\text{life})} \quad (5)$$

and

$$P(\text{context} \mid \text{no life}) = \frac{P(\text{no life} \mid \text{context})P(\text{context})}{P(\text{no life})}. \quad (6)$$

We can substitute the expressions of Eqs. 5 and 6 into Eq. 3. Then, $P(\text{context})$, $P(\text{life})$, and $P(\text{no life})$ terms will cancel out to give the expression we seek: the conditional probability of the hypothesis of life being present given spectral or photometric data (with possible biosignatures) and the context of the extrasolar system. The expression is as follows:

$$P(\text{life} \mid D, C) =$$
$$\frac{P(D \mid C, \text{life})P(\text{life} \mid C)}{P(D \mid C, \text{life})P(\text{life} \mid C) + P(D \mid C, \text{no life})P(\text{no life} \mid C)}. \quad (7)$$

Here, to make the expression less unwieldy, we have substituted $D$ for "data" and $C$ for "context."

In Eq. 7, the probability $P(\text{life} \mid D, C)$ on the left-hand side is the ''posterior probability'' that we seek, which depends on a weighted assessment of the hypothesis of life's presence given the chance of the data $D$ occurring in the given context $C$ of exoplanet system properties, as expressed on the right-hand side of the equation. In the numerator and denominator, $P(D \mid C, \text{life})$ is the conditional probability—the ''likelihood''—of the data $D$ occurring in the given context $C$ of an exoplanet if life is present. The likelihood of the data $D$ occurring when life is not present on the exoplanet is the probability $P(D \mid C, \text{no life})$. Importantly, $P(D \mid C, \text{no life})$ incorporates the idea of an abiotic *false positive* detection of life.

We shall use the term ''context-type'' or *C*-type for an exoplanet's context. In the general case, an exoplanet context is a continuous function, and so a limitless number of possible parameter permutations define that context. What we mean by ''*C*-type'' is a discrete class of exoplanet based on how a researcher chooses to group exoplanets. For example, a *C*-type could be Earth-size and -mass exoplanets in the habitable zone (HZ) around G-stars. How these particular classes are chosen depends on what someone wants to quantify. Later, we give an example wherein the *C*-type of an exoplanet is discretized based solely on the spectral type of the parent star. In this case, a researcher could be attempting to quantify the prevalence of life (or habitability) around G-stars versus K-stars, and so on, and so exoplanets would be grouped by spectral type of the host star. One could imagine other bins (or *C*-types) based on specification of planet size, mass, surface emission, and so on.

In Eq. 7, we also have a ''prior'' probability $P(\text{life} \mid C)$, which is the prior estimated chance of life being present on an exoplanet given all current scientific knowledge about the *C*-type of an exoplanet at the time of its assessment. Similarly, the ''prior'' $P(\text{no life} \mid C)$ is the estimated chance of no life being present on an exoplanet, given all current knowledge of its context. Because we assume that a biosphere is either present on an exoplanet or absent, it follows that these two are priors are related by

$$P(\text{no life} \mid C) = 1 - P(\text{life} \mid C). \tag{8}$$

The ''priors'' of $P(\text{life} \mid C)$ and $P(\text{no life} \mid C)$ combine two strains of scientific thought. The first is prior knowledge of habitability and the second is the estimated chance of life emerging on a habitable exoplanet (which is discussed further later). Regarding habitability, the ubiquitous concept of the HZ is built upon the assumption that $P(\text{life} \mid C)$ is higher where the context is an exoplanet orbiting within a certain range of star–planet distances (Kasting *et al.*, 1993; Kasting, 1997; Kopparapu *et al.*, 2013). Consequently, the HZ should be considered as a probability density function (Zsom, 2015). Physically, the HZ is the region around a star where prior knowledge suggests that an exoplanet can maintain liquid water on its surface. The conventional limits of the HZ are calculated with an Earth-like climate assumed to arise from a $CO_2$–$H_2O$-rich atmosphere. This assumption has been questioned because thick $H_2$-rich atmospheres can provide warming far beyond the conventional HZ, even in interstellar space (Stevenson, 1999; Pierrehumbert and Gaidos, 2011; Seager, 2013). However, the conventional HZ provides a guide based on real prior knowledge about $CO_2$–$H_2O$-rich planets from a real world—the Earth—whereas habitable $H_2$-rich worlds remain hypothetical speculations.

As already mentioned, life's presence on an exoplanet tells us something about the exoplanet's context. For example, if a planet has life, we would infer that the planet is not orbiting so close to its host star that life would be incinerated. Furthermore, some environmental contextual parameters on the planet may be modulated by life, for example, the bulk atmospheric composition and resulting climate. On Earth, the biosphere modulates all the major atmospheric gases, such as $O_2$ and $N_2$, and most trace gases with the exception of noble gases (*e.g.*, Catling and Kasting, 2017). Consequently, the distinction between inhabitation and habitability can become blurred (Goldblatt, 2016). Thus, the correct prior for assessing biosignatures is $P(\text{life} \mid C)$, which captures the intertwining of life and environment rather than $P(\text{life})$. We consider the latter to be a probability that one derives in the future from the results of observational statistics of noninhabitation or inhabitation of exoplanets. We illustrate this point later.

To better understand Eq. 7, consider a simple example applied to single exoplanet. Suppose the prior, $P(\text{life} \mid C)$, is appreciable, for example, 0.5. This prior is a fifty-fifty chance that a planet with a certain context *C* (such as the HZ of a G-star) actually has life. From Eq. 8, the prior, $P(\text{no life} \mid C)$ is 0.5. In this case, a possible detection of biosignatures will tend to be influenced by the estimated likelihood $P(D \mid C, \text{life})$, that is, the estimate of the conditional probability of the occurrence of the data $D$ given the context $C$ and hypothesis of life being present. Suppose an Earth-sized exoplanet with evidence of a liquid ocean is found to have an $O_2$-rich atmosphere. Under these circumstances, also suppose that the best models of the exoplanet suggest $P(D \mid C, \text{life}) = 0.80$ and $P(D \mid C, \text{no life}) = 0.25$. One can think of the latter as the probability of the data $D$ representing a ''false positive.'' Putting these numbers into Eq. 7, the posterior probability of life being present on the exoplanet will be

$$P(\text{life} \mid D, C) = \frac{0.8 \times 0.5}{(0.8 \times 0.5) + (0.25 \times 0.5)} = 0.76 \equiv 76\%. \tag{9}$$

Alternatively, if the prior for the presence of life given the context is very small, that is, $P(\text{life} \mid C) << 1$ and so the prior for ''no life'' is large $P(\text{no life} \mid C) \to 1$, then Eq. 7 suggests that $P(\text{life} \mid D, C)$ would tend toward $P(\text{life} \mid C) \times [P(D \mid C, \text{life})/P(D \mid C, \text{no life})]$, that is, the value of ''prior'' probability for life (given the context) times the ratio of the likelihoods of the data with and without life. This posterior probability would be low and dominated by the small prior, $P(\text{life} \mid C)$, unless the probability of a false positive is very small. In the former case, the ''base rate'' of uninhabited *C*-type exoplanets, which is reflected by the small prior $P(\text{life} \mid C)$, would far outweigh the number of inhabited *C*-type exoplanets. Then false identifications of life are mathematically anticipated given that an estimated ratio $P(D \mid C, \text{life})/P(D \mid C, \text{no life})$ may be sizeable if models optimistically overestimate the extent that life might be responsible for the data. This Bayesian insight cautions us that exoplanet data might be overoptimistically interpreted to indicate the presence of life, if life is actually very rare in the galaxy.

In reality, of course, the probabilities shown in Figure 1 will not be single numbers but functions of numerous parameters. As a result, the posterior probability that we seek, $P(\text{life} \mid D, C)$, will be a more nuanced function of multiparameter space. However, some variables will be unimportant. When we calculate $P(\text{life} \mid D, C)$ by integration, that is, for all possible variables weighted by their probability of occurrence, we expect that a reduced set of those variables—the so-called marginal variables in Bayesian statistics—will be important, whereas nuisance variables can be discarded, that is, "marginalized out."

It is informative to think about how the various terms in Eq. 7 are related to the well-known Drake equation. From that comparison, we can then consider what the terms mean for developing a practical framework of assessing biosignatures on exoplanets and assigning a probability to the presence of life.

The probability $P(\text{life} \mid \text{context})$ shown in Figure 1 refers to a particular exoplanet under observation but is related to the term $f_l$ in the Drake equation that applies to an ensemble of exoplanets (Des Marais, 2015). Specifically, in the Drake equation, $f_l$ is defined as "the fraction of... [habitable] planets on which life actually develops" (Drake, 1965). A related quantity can be derived from our formalism, which is the "base rate" frequency of occurrence of inhabited exoplanets around all main sequence stars in the galaxy, which we denote as $\langle P(\text{life})\rangle$. Rather than a single exoplanet, this derived quantity would require multiple nondetections of life or even detections of multiple biospheres to be quantified, each observation causing an adjustment of priors.

The calculation of $\langle P(\text{life})\rangle$ is related to the prior, $P(\text{life} \mid C)$. The quantity $\langle P(\text{life})\rangle$ is calculated by integrating over the probability density of all $C$-type systems as follows:

$$\langle P(\text{life})\rangle = \sum_C P(\text{life} \mid C) P(C). \tag{10}$$

To give an illustrative example, the $C$-type terms could be main sequence stellar classes of exoplanet host stars. In that case, we could expand Eq. 10 as

$$\begin{aligned}\langle P(\text{life})\rangle = &\, P(\text{life} \mid \text{M-stars}) P(\text{M-stars}) \\ &+ P(\text{life} \mid \text{K-stars}) P(\text{K-stars}) \\ &+ P(\text{life} \mid \text{G-stars}) P(\text{G-stars}) + \ldots\end{aligned} \tag{11}$$

Here, $P(\text{life} \mid \text{M-stars})$, *etc.* refer to the prior probability distributions of life occurring in M-star systems, and so on, and the probabilities $P(\text{M-stars})$, $P(\text{K-stars})$, *etc.* are the observed occurrence probabilities in our galaxy of main sequence stellar types, which sum to unity: 0.76 for M-stars, 0.12 for K-stars, 0.076 for G-stars, 0.03 for F-stars, 0.006 for A-stars, 0.0013 for B-stars, and $\sim 10^{-5}$ for O-stars (Ledrew, 2001). The prior conditional probabilities $P(\text{life} \mid \text{M-stars})$, *etc.* are currently unknown, except we know that $P(\text{life} \mid \text{G-stars})$ is nonzero because life exists on Earth. Most people would also consider it reasonable to assume that $P(\text{life} \mid \text{O-stars})$ is negligible because O-stars have main sequence lifetimes that range from less than one to a few million years (Weidner and Vink, 2010).

Because $\langle P(\text{life})\rangle$ in Eq. 11 covers all main sequence stellar types and exoplanets, and is for life present only at the current observed time, it will be less than $f_l$ in the Drake equation, which refers to life arising on habitable planets at any time. But we could consider a restricted case to relate our terms to $f_l$. To do this, we consider only four $C$-type systems, habitable exoplanets in M-, K-, G-, and F-star systems rather than all stellar systems, because this restriction is usually assumed in evaluating the Drake equation. Then, we also redefine our prior conditional probability distributions to be terms such as $P(\text{life} \mid \text{M-plan}_{\text{hab}})$, meaning the probability of life ever occurring during the lifetime of the planet given the context of a habitable M-star planet. In addition, we weigh by the probability of the specific $C$-type systems: $P(\text{M-plan}_{\text{hab}})$, $P(\text{K-plan}_{\text{hab}})$, *etc.*, meaning the fraction of habitable planets that orbit M-stars, K-stars, *etc.* With such revised definitions, the $f_l$ term in the Drake equation would be analogous to Eq. 11, that is

$$\begin{aligned}f_l = &\, P(\text{life} \mid \text{M-plan}_{\text{hab}}) P(\text{M-plan}_{\text{hab}}) \\ &+ P(\text{life} \mid \text{K-plan}_{\text{hab}}) P(\text{K-plan}_{\text{hab}}) \\ &+ P(\text{life} \mid \text{G-plan}_{\text{hab}}) P(\text{G-plan}_{\text{hab}}) \\ &+ P(\text{life} \mid \text{F-plan}_{\text{hab}}) P(\text{F-plan}_{\text{hab}})\end{aligned} \tag{12}$$

The estimated "base-rate" frequency of occurrence of life on $C$-type exoplanets, $P(\text{life} \mid C)$, deserves some more discussion. Recall that this prior is our best scientific estimate of the occurrence of life on a $C$-type exoplanet given all current knowledge. Thus, this estimate would be significantly improved if we actually detect any life elsewhere or found environments on exoplanets to support the HZ concept. Then our prior statistical information about the presence of life would expand beyond just the Earth and improve the estimate of the true frequency that life occurs on a $C$-type exoplanet.

Because extraterrestrial life has not yet been detected, the present state of scientific knowledge about the *a priori* probability of life occurring in the $C$-type that is an Earth-like, habitable exoplanet is based on two broad approaches: (1) consideration of the origin and persistence of life on a planet from laboratory and theoretical studies and (2) the fact that detectable life exists on Earth, a habitable planet within the Sun's HZ.

Unfortunately, in the first approach about how easily life originates on a habitable exoplanet $C$-type, conflicting views persist. Some argue that life readily emerges from a habitable environment because biochemistry naturally emerges from geochemical reactions on an Earth-like planet (Smith and Morowitz, 2016). Others argue that the origin of life is extremely improbable even if we reran the clock on Earth (Monod, 1971; Koonin, 2012). The opinions in this first approach range from $P(\text{life} \mid \text{Earth-like})$ being nearly unity to being vanishingly small. However, $P(\text{life} \mid \text{Earth-like})$ is *not* a totally unknown probability where, following Laplace, we would have to assume it is equally likely to have any value from 0 to 1. The Earth is inhabited. Consequently, $P(\text{life} \mid \text{Earth-like}) > 0$, which is an aspect of empirical astronomy to which we now turn. Nonetheless, given the present stage of understanding and lack of data, the possibilities for $P(\text{life} \mid \text{Earth-like})$ range from life being very common to very rare, as has been much discussed in the literature (*e.g.*, Carter, 1983; Spiegel and Turner, 2012; Scharf and Cronin, 2016; Walker, 2017).

Earth and Mars are two rocky planets currently in the HZ of the Sun (Kopparapu *et al.*, 2013). No data from Mars conclusively show that it ever had a biosphere (Klein, 1998; Cottin *et al.*, 2017), although this situation could change if future Mars exploration uncovers evidence of life. Another consideration is that if Mars were truly more Earth-like, that is, bigger, it may have been able to hold on to its volatiles and recycle them through active volcanism (Catling and Kasting, 2017, Chap. 12), and perhaps life would be present on such a different Mars today. Noting those caveats, in the absence of any other information, if we found an Earth-like planet in the HZ of a Sun-like star, a starting point—based on empirical data of our current solar system—might be to assume a prior of $P(\text{life} | \text{Earth-like}) = 0.5$ (noting that at a different time, Venus, may have been habitable $\sim$4 billion years ago, at least in principle, when the Sun was fainter). One can easily criticize this rough approach. But with the current absence of knowledge, it is a practical point of departure, and, as we explain later, if future data from exoplanets become increasingly difficult to explain without invoking the presence of life, it turns out that the exact value for the assumed ''prior'' does not matter.

The path forward for assessing whether exoplanets have life is a path somewhat similar to that followed by Borucki *et al.* (1996) in conceiving the *Kepler* mission to determine the number of Earth-sized planets in the Milky Way in the face of a great range of opinion about whether any stars even had such planets. Direct observation of the Cosmos is the only sure path.

To be successful in assigning probabilities to the future detection of life on an exoplanet, astrobiology research must constrain the conditional probabilities $P(D | C, \text{life})$ and $P(D | C, \text{no life})$ (Fig. 1). These likelihoods do not necessarily sum to unity, as illustrated in the previous illustrative example. To constrain $P(D | C, \text{life})$, a biosignature research framework must explore how exoplanet spectra or photometry data can arise from various theoretical permutations of inhabited Earth-like planets with different biospheres (*e.g.*, oxygenic photosynthetic, anoxic photosynthetic, or nonphotosynthetic) under the oxidizing or reducing atmospheres that we expect are the sequence of the chemical evolution of atmospheres on rocky worlds, and will be part of the data under assessment. Similarly, to constrain $P(D | C, \text{no life})$, we must simulate the spectra or photometry of uninhabited planets, including planets that could plausibly produce false positives.

To estimate the likelihoods, we must have relevant contextual data, such as stellar parameters and physical properties of the exoplanet, as we discuss in sections 3.1 and 3.2. Moreover, to constrain $P(D | C, \text{no life})$, the contextual information needs to be used in ''Exo-Earth System'' models to generate synthetic exoplanet spectra or photometric data (Fig. 1).

''Exo-Earth System'' models simulate many parts of an Earth-like planet, such as the interior (*e.g.*, mantle thermal evolution), ocean, biosphere, and atmosphere. The biosphere, for example, could be implemented as fluxes of biogenic gases of interest that are parameterized to depend on temperature, nutrients, and biological productivity, constrained by chemical stoichiometry and redox balance (*e.g.*, Claire *et al.*, 2006; Gebauer *et al.*, 2017). Alternatively, ''Exo-Earth System'' models could be used in an inverse fit of model parameters to observational data. In any case, such models are essential tools that need to be developed for a framework of assessing exoplanet biosignatures, as discussed later in section 3.3.

The importance of false positives is highlighted in our proposed framework (Eq. 7 and Fig. 1). In the future, if data from an exoplanet become more and more difficult to explain by abiotic explanations (*i.e.*, $P(D | C, \text{no life})$ approaches zero), then the second term in the denominator of Eq. 7 will become negligible. Hence, the probability of life on an exoplanet with data suggestive of biosignatures will converge to 1 [$P(\text{life} | D, C) \rightarrow 1$ in Eq. 7] regardless of the exact choice of the ''prior'' $P(\text{life} | C)$ and irrespective of the exact computed likelihood $P(D | C, \text{life})$. This Bayesian insight reveals how critical it is to evaluate the plausibility of a false positive detection of life. If proposed false positive scenarios prove to be less realistic than some hypothesize, detecting life from Earth-like exoplanet biosignatures becomes mathematically more favorable.

## 3. Input Components in a Framework of Observing Exoplanet Biosignatures

The Bayesian framework illustrated in Figure 1 begs the question of which exoplanet parameters we need to measure to best assess biosignatures. To this end, some practical steps are shown in a four-component procedure of observation and interpretation in Figure 2. Each component is an observational and/or analytical procedure intended to increase confidence and reduce uncertainty in a potential biosignature. For practical reasons—the schedule of new telescopic observations and when new instruments see first light—measurement components may not follow the idealized sequence represented by yellow arrows in Figure 1. For example, in the case of next-generation telescopes, ''external'' exoplanet contextual properties shown in Figure 1 (*e.g.*, exoplanet mass) might be determined after an exoplanet spectrum has been acquired.

In any case, the first two components shown in Figure 2 are needed to gather astrophysical and planetary information (the ''context'' box of Fig. 1) for a probabilistic assessment of life on an exoplanet, and may also include some elements of the data box shown in Figure 1. The first component is to characterize ''external'' properties of an exoplanetary system, including the properties of the host star, the orbital and physical properties of the system, and the mass and radius of a target exoplanet. These properties can be fed into Exo-Earth models to simulate exoplanet data. The second component involves characterization of the key ''internal'' properties of the target exoplanet, ideally including its bulk atmospheric composition, global mean climate, and surface material properties. Properties that are considered independent of life can be fed into Exo-Earth models also. Otherwise, properties would need to be considered as part of the biosignature data rather than contextual data.

The third and fourth components shown in Figure 2 gather and examine the spectral or photometric data that contain potential biosignatures and include the information that lies within the ''data'' box of Figure 1. The third component is the explicit search for biosignatures in the reflected spectrum, transmission spectrum, emission spectrum, or photometry of an exoplanet. These data are those that Exo-Earth models attempt to simulate under scenarios of life being present or absent, which can be done by using either forward or inverse (retrieval) methods.

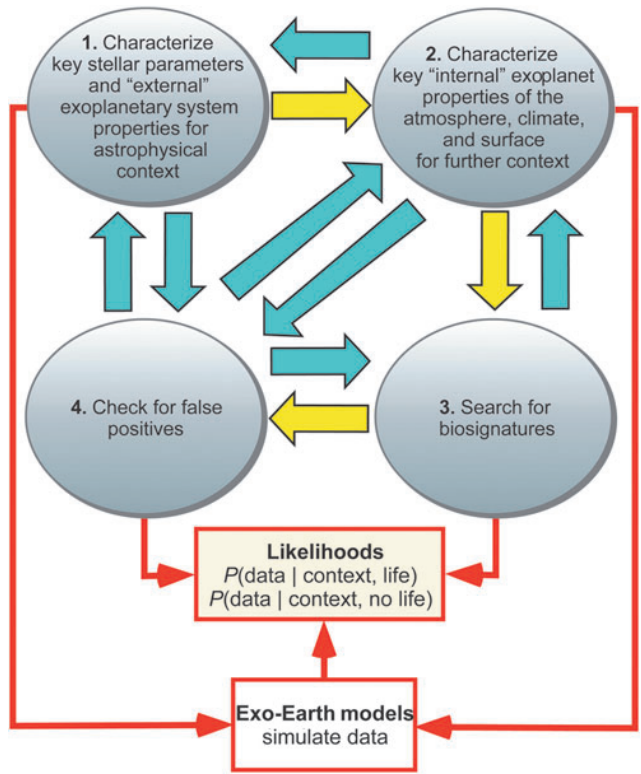

**FIG. 2.** Four components to assess whether potential biosignatures are best explained by the presence of life. The numbered order and yellow arrows indicate information that an idealized observational strategy would gather sequentially, although in reality the order will likely be different. The blue arrows indicate how practical observation and analysis would require iteration to increase confidence in biosignature acceptance. Alternatively, following the blue arrows could aid in identification of a false positive. These four components, combined with Exo-Earth models (see text), would help to constrain the likelihoods of the biosignature data occurring with and without life: $P(\text{data}\,|\,\text{context, life})$ and $P(\text{data}\,|\,\text{context, no life})$, respectively.

The fourth component is a procedure to distinguish a truly biogenic signal from all conceivable false positives. This component requires Exo-Earth modeling to test potential false positive scenarios and generates $P(D\,|\,C, \text{no life})$. This procedure likely requires iteration with previous components. The result of the four components is to estimate the likelihoods given in the cream-colored box shown in Figure 2 that are required in the overall scheme of Figure 1.

In the following subsections, we break down each of the four procedural components of this framework into more detail. We also comment on the necessity to model an Earth-like planet and so simulate the spectral or photometric data that bear upon the possible presence of life.

### 3.1. Component 1: stellar properties and "external" properties of the exoplanetary system

A potentially inhabited exoplanet is embedded within the environment of its host star, so knowledge of the properties of the host star is critical for understanding exoplanet habitability. In addition, some "external" properties of the exoplanetary system are either essential or desirable to know for Exo-Earth models of Figure 1. Table 1 lists potential properties of a star that would be advantageous to determine, as well as key exoplanetary system parameters. These properties and parameters are relevant for determining whether the exoplanet is habitable or inhabited in ways discussed next and in section 3.2.

*3.1.1. Stellar parameters.* Knowing the age of the star and current luminosity provides context for assessing the potential evolution of an exoplanet's atmosphere and possible life. First, main sequence stars brighten with age, and this affects an exoplanet's atmosphere and habitability (Shklovskii and Sagan, 1966; Sagan and Mullen, 1972). As a result, the HZ lifetime depends on stellar evolution (Rushby *et al.*, 2013). Second, a very young planet is unlikely to have developed life that had time to evolve the complex biochemistry necessary to substantially alter its environment, and hence produce a detectable biosignature. For example, in the case of Earth, $O_2$ was not detectable in our atmosphere until about halfway through Earth's history (Lyons *et al.*, 2014). It is likely that this was at least, in part, due to the geochemical steps needed before $O_2$ could accumulate in the atmosphere. Some have argued, on the grounds of redox balance, that the long timescale for the Earth's atmospheric oxygenation was a result of a slow global "redox titration" of materials in the Earth's crust (Catling and Claire, 2005; Catling *et al.*, 2005; Zahnle and Catling, 2014), which some hypotheses relate to gradual tectonic and/or mantle evolution (Holland, 2009; Kasting, 2013). Although stellar ages are sometimes poorly known, the upcoming planetary transits

TABLE 1. DESIRABLE KEY PROPERTIES TO KNOW ABOUT THE HOST STAR AND EXOPLANETARY SYSTEM

| *Stellar properties* | *Exoplanetary system properties* |
|---|---|
| Age | Physical exoplanet properties |
| Spectral type (including effective temperature) |   Mass and radius: hence mean density and bulk composition |
| Stellar luminosity |   Presence of a surface |
| Panchromatic spectrum, particularly the UV flux | |
| Activity | Orbit and spin parameters |
| Rotation rate (related to age and X-ray flux, which correlates with emission of strong UV lines) |   Orbital eccentricity |
| |   Rotation rate |
| Elemental composition |   Obliquity |
| Whether in a multiple star system, *e.g.*, a binary star system | Other planets in the system, and their physical and orbital properties (*e.g.*, to determine resonances) |

and oscillations of stars mission will improve estimates (Rauer *et al.*, 2014).

The spectral type and effective temperature of a star are essential parameters for estimating the stellar flux on an exoplanet, which is a key parameter for the planetary climate and habitability (*e.g.*, Pierrehumbert, 2010; Kopparapu *et al.*, 2013; Catling and Kasting, 2017) (see section 3.2).

The panchromatic spectrum, including the UV flux, provides an essential input for models of the atmospheric chemistry, climate, and atmospheric escape of exoplanets (Linsky and Güdel, 2015). The measurement of UV must be short wavelength enough to include Lyman-$\alpha$ output at 121 nm, which causes much of the photolysis of molecules such as $CH_4$ and $H_2O$. The spectral energy distribution (SED) is needed to understand atmospheric chemistry because it affects the reactions of key molecules of interest on a habitable exoplanet, such as $CO_2$, $CH_4$, $O_2$, and $H_2O$ (Segura *et al.*, 2003; Grenfell *et al.*, 2007; Rugheimer *et al.*, 2013). In particular, the abundances of potential biosignature molecules such as $CH_4$ in oxic atmospheres are strongly affected by the magnitude of a star's near-UV flux <310 nm because this wavelength region generates the OH (hydroxyl) radical, which is a key oxidizing species (Segura *et al.*, 2005). Indeed, the most common stars, M dwarfs, have a lower flux in the near-UV than the Sun, so that for the same biogenic $CH_4$ flux as the Earth, the $CH_4$ abundance reaches a higher and more detectable level depending upon the exact SED (Segura *et al.*, 2005; Rugheimer *et al.*, 2015). The SED is also necessary to run photochemical models that disentangle possible gaseous biosignatures from abiotic false positives, as explored for $O_2$ and $O_3$ (Domagal-Goldman *et al.*, 2014; Tian *et al.*, 2014; Gao *et al.*, 2015; Harman *et al.*, 2015; Meadows, 2017) and that would be expressed as $P(D\,|\,C$, no life) in our framework of Figure 1.

The SED also directly affects the HZ, which is a key part of the context-dependent prior, [$P$(life $|\,C$) or $P$(no life $|\,C$)] shown in Figure 1. Near the inner edge of the HZ, a planet orbiting a cooler star has a lower albedo in the near-infrared (NIR) and reaches the "moist greenhouse" limit of significant water loss at a lower incident flux than a warmer star due to NIR absorption by $H_2O$ and $CO_2$ and weaker Rayleigh scattering (*e.g.*, Catling and Kasting, 2017, pp 427–430; Kopparapu *et al.*, 2013). Near the outer edge of the HZ, the ice-albedo feedback is weaker for a cooler star because sea ice and snow are less reflective in the NIR (Joshi and Haberle, 2012; Shields *et al.*, 2013).

Stellar activity is dominated by the stellar magnetic field evolution, which affects flare occurrence, the intensity of a stellar wind, and the emission at UV and X-ray wavelengths. For HZ planets around M-stars, compression of planetary magnetospheres by intense stellar winds might make a planetary surface susceptible to cosmic ray exposure (Griessmeier *et al.*, 2005). Consequently, knowledge of stellar activity is desirable since it may affect our prior, $P$(life $|\,C$). It has also been proposed that magnetospheric compression due to strong coronal mass ejections from young stars allows energetic particles to induce consequential upper atmospheric chemistry. The resulting reactions may produce gases such as $N_2O$ (Airapetian *et al.*, 2016). Modeled upper atmosphere $N_2O$ at part per million levels could present a false positive for $N_2O$, which is a biogenic gas on Earth, that is, affecting the likelihood of data, $P(D\,|\,C$, no life). In contrast, modeled ground-level concentrations of $N_2O$ are <<1 ppbv and so would produce negligible greenhouse warming with little impact on habitability given that Earth's modern concentration of $N_2O$ of $\sim$0.32 ppmv (parts per million by volume) produces <1 K of warming (Schmidt *et al.*, 2010).

Stellar activity is related to the rotation rate of a star (Pallavicini *et al.*, 1981; Ribas *et al.*, 2005), so knowledge of the stellar rotation rate would help characterize the shortwave emission of a host star, particularly the large Lyman-$\alpha$ output at 121 nm (Linsky *et al.*, 2013), which produces photons that drive photolysis reactions in planetary atmospheres, and feeds into estimated data likelihoods shown in Figure 1. Stellar rotation can be measured in various ways (Slettebak, 1985; Rozelot and Neiner, 2009): photometric tracking of features on the star (such as starspots), Doppler shift in the stellar spectrum, using a star's oblateness (van Belle, 2012), or using the Rossiter effect that distorts the radial velocity curve in rotating binary star systems.

The elemental composition of the parent star, although not essential for biosignature interpretation *per se*, is desirable to help understand the potential bulk composition of the nebula that gave rise to the exoplanet and its bulk composition. Elemental composition gives metallicity—the proportion of elements other than hydrogen or helium expressed on a $\log_{10}$ scale relative to the Sun. Metallicity is a parameter commonly used in astronomy, where solar is 0, metal poor <0, and metal-rich >0. A correlation between gas giant planet occurrence and stellar metallicity exists (Fischer and Valenti, 2005; Johnson *et al.*, 2010) and metallicity differences also appear to be linked to the occurrence of rocky planets and small gas-rich planets (Wang and Fischer, 2015).

At present, apart from the bulk density, the only insight into the bulk composition of an exoplanet comes from indirect inference from the host star's elemental composition through the spectrum of its photosphere (Gaidos, 2015; Dorn *et al.*, 2017). An exoplanet's bulk composition is potentially useful for modeling a planet's possible geological evolution, which would affect our data likelihoods shown in Figure 1. For example, a C/O ratio >1 might preclude plate tectonics because SiC rocks might form rather than silicates (Kuchner and Seager, 2005).

Finally, about one-third of stars in the Milky Way are in binary or multiple systems (Lada, 2006; Duchene and Kraus, 2013), and the long-term habitability of exoplanets may be affected by the presence of a nearby star (David *et al.*, 2003). A key concern is whether the orbits of HZ planets in multiple star systems are stable on sufficiently long timescales to be compatible with an origin and evolution of life, which may take $10^8$–$10^9$ years. In addition, multistar systems can host the remnants of an evolved star (*e.g.*, a white dwarf), which would indicate that the system, at some point, endured dramatic changes during the post-main sequence evolution of the evolved companion. Consequently, it would be desirable to know whether the exoplanet host star is part of a multiple star system to understand whether the exoplanet may have been subject to a history that makes it a less viable candidate for inhabitation. Also, sometimes the age of a star is determined from the better known age of a companion, for example, Proxima Centauri's $\sim$4.8 Ga age is estimated from that of alpha-Centauri (Bazot *et al.*, 2016).

3.1.2. *Planetary system parameters.* Among basic physical properties, the exoplanet mass and radius provide the mean density of the exoplanet and the gravitational acceleration at the surface, $g$, which are very useful contextual parameters for the likelihoods and priors shown in Figure 1. The mean density sets overall constraints on rockiness and the possibilities for the size of an iron-rich core. Consequently, an Earth-like mean density would lend confidence that an exoplanet is truly rocky and not a migrated icy body or a planet with a large gas envelope.

An accurate internal structure model of an exoplanet, of course, remains undefined without a moment of inertia. Some proposals have been made for determining the moment of inertia of exoplanets from the influence of a gravitational quadrupole field and tidal dissipation characteristics in the case of certain orbital architectures of exoplanetary systems (Batygin *et al.*, 2009; Mardling, 2010; Batygin and Laughlin, 2011; Becker and Batygin, 2013).

The gravitational acceleration, $g$, of an exoplanet is necessary for understanding a variety of atmospheric parameters, including the scale height and lapse rate (see section 3.3.2). These parameters are needed to accurately estimate the abundances of gases in an exoplanet's atmosphere from retrievals and a possible inference of the surface temperature of the planet from calculated greenhouse warming if surface temperature is not directly observable from thermal emission.

The value of $g$ also affects spectral interpretations. For example, Rayleigh scattering depends on column number density, which depends on $g$. Transit spectra are sensitive to the pressure scale height of an atmosphere, which is an inverse function of $g$ (*e.g.*, Seager, 2010, pp 44–45).

Knowledge of the rotation rate of an exoplanet is highly desirable for a basic model of climate and habitability, and is context that would affect both likelihoods and priors in shown Figure 1. The rotation rate controls both the temporal distribution of irradiation from the host star and the dynamical regime of an exoplanet's atmosphere (Showman *et al.*, 2013). Essentially, slowly rotating planets tend to be in ''all tropics'' dynamical regimes with equatorial superrotation. Fast rotation (typically with rotation periods from tens of hours down to a few hours) causes a climate system to become more structured with latitude: smaller and more numerous Hadley, shorter eddy length scales, and more jets (Kaspi and Showman, 2015). If the rotation period becomes comparable to or greater than the radiative relaxation time, strong day–night contrasts emerge, generating thick dayside clouds that might extend habitability closer to the host star than the traditional inner edge of the HZ (Yang *et al.*, 2014; Kopparapu *et al.*, 2016; Way *et al.*, 2016).

In principle, the rotation rate can be inferred in several ways: (1) from brightness or color variations as a planet rotates (including the effect of continental or persistent cloud patterns) (Ford *et al.*, 2001; Palle *et al.*, 2008; Cowan *et al.*, 2009; Oakley and Cash, 2009; Livengood *et al.*, 2011); (2) from the Doppler shift of absorption features during ingress and egress of a transiting exoplanet (Spiegel *et al.*, 2007; Brogi *et al.*, 2016); or (3) from comparing thermal phase curve with rotation-dependent models (Rauscher and Kempton, 2014).

Knowledge of an exoplanet's obliquity (axial tilt) would give a more complete picture of habitability through the effect of seasonality on heat transport around the planet. A large obliquity may potentially allow for large temperature excursions on a planet's surface, which would affect habitability (Williams *et al.*, 1996). Alternatively, obliquity with high frequency variations could suppress ice-albedo feedback and extend the HZ outwards (Armstrong *et al.*, 2014). However, determining exoplanet obliquity would require analyses of temporal color changes at different orbital phase angles (Kawahara and Fujii, 2010; Kawahara, 2016).

The orbital eccentricity is also essential for a model of the possible atmospheric evolutionary history, the seasonal climatic variations on an exoplanet, and potential planet–planet dynamical interactions. Although climate variations alone due to eccentricity may not compromise habitability (Williams and Pollard, 2002; Dressing *et al.*, 2010; Kane and Gelino, 2012; Way and Georgakarakos, 2017) except perhaps in extreme cases (Bolmont *et al.*, 2016), planets with very high eccentricity could experience large tidal heating that could desiccate a planet through water loss (Barnes *et al.*, 2013). In contrast, tidal heating could also prevent global snowball states (Reynolds *et al.*, 1987; Barnes *et al.*, 2008).

Knowing about other planets and their properties in an extrasolar system could be important for deducing the formation history of the star–planet system, the possible orbital evolutionary path of the exoplanet of interest, and providing information that bears on climate stability. Orbital elements and obliquity of rocky planets are affected by neighboring planets and moons (*e.g.*, the Milankovitch cycles on Earth). Indeed, in binary star systems, climate cycling can be extremely rapid (Forgan, 2016). Planet–planet mean orbital motion resonances might also affect HZ planets, as in the system of seven planets around the star TRAPPIST-1 (Gillon *et al.*, 2017). Such resonances can cause tidal heating and be related to a history of planet migration, whereby planets may have moved into or out of a HZ over time. A planet could also appear habitable but might have large, biologically harmful variations of eccentricity or inclination perhaps forced by a highly inclined planetary neighbor in the Lidov–Kozai mechanism of three bodies where inner binary planets are affected by a third faraway companion (Naoz, 2016).

A full inventory of an extrasolar system would include asteroid or comet populations. IR, submillimeter, and millimeter observations are a possible means of constraining such debris and materials (*e.g.*, Anglada *et al.*, 2017). Combined with knowledge of other perturbing bodies, knowledge of small bodies might allow estimates of impact rates, giving insight into the evolutionary stage of the planetary system and volatile delivery or erosion (*e.g.*, Zahnle and Catling, 2017).

### 3.2. Component 2: the ''internal'' properties of the exoplanet atmosphere, climate, and surface

A planet inside a HZ is not necessarily habitable. Stellar and exoplanetary system external properties described for ''component 1'' provide context for understanding an exoplanet's habitability, but direct measurements concerning an exoplanet's ''internal'' properties are also needed to assess habitability, including bulk atmospheric composition and structure, climate, and surface state (Table 2). Such data feed into the scheme of Figure 1 in various ways, for example, the prior, $P\,(\mathrm{life}\,|\,C)$, would increase if we knew that an exoplanet surface has an ocean. Alternatively, atmospheric properties (*e.g.*, surface pressure) would be useful for generating likelihoods of the data shown in Figure 1.

Table 2. Desirable "Internal" General Properties of an Atmosphere and Surface to Know About a Potentially Habitable Exoplanet

| Climatic and surface properties | Atmospheric properties |
|---|---|
| Average surface emission temperature, $T_{\text{surfe}}$ | Bulk composition and redox state |
| Bond albedo, $A_B$ | Surface barometric pressure |
| Effective temperature, $T_{\text{eff}}$ | Clouds or haze composition and structure |
| Composition of the surface phase: | Vertical structure |
|   Liquid water or ice | Mixing ratios of key trace species: |
|   Silicates |   $H_2O$ from evaporation |
| |   Volcanic gases |
| |     (e.g., $SO_2$ or $H_2S$) |
| |   Greenhouse gases (see Table 3) |

Measurements such as the surface temperature or the detection of liquid water on the surface would provide direct confirmation that an exoplanet is habitable (Robinson, 2018), after the conventional definition of a habitable planet as one with a solid surface that has an extensive covering of liquid water (Kasting *et al.*, 1993). Other data given in Table 2 would increase confidence that the exoplanet is Earth-like in detail rather than just similar in size and mass, like Venus.

3.2.1. *Climatic and surface parameters and properties.* Knowing whether the global average surface emission temperature $T_{\text{surfe}}$ of an exoplanet is compatible with liquid water is highly desirable and possibly essential to be highly confident of the detection of life on an exoplanet from biosignatures (section 3.3), and is context that affects both our prior and data likelihoods shown in Figure 1. On a planet with a thick atmosphere and surface water, the value of $T_{\text{surfe}}$ must also be less than some upper limit for biomolecule stability. On Earth, deep-sea hyperthermophilic microbes can grow at 121°C and survive up to 130°C (Kashefi and Lovley, 2003). But many biomolecules rapidly degrade >150°C (White, 1984; Somero, 1995) irrespective of the influence of high pressure, which may be stabilizing or destabilizing (Lang, 1986; Bains *et al.*, 2015). Thus, a conservative upper limit on $T_{\text{surfe}}$ for a habitable planet is $T_{\text{surfe}} < 425$ K. A surface temperature 647 K—water's critical temperature at which water can only exist as vapor—is the likely firm physical limit for habitability.

In principle, $T_{\text{surfe}}$ could be inferred directly if an exoplanet is observed over a wavelength region in which its surface emits thermal radiation to space through a nearly transparent window where the atmosphere does not absorb. For example, the Earth's surface temperature can be observed from space through an atmospheric window at wavelengths between 8 and 12 μm. However, thick hazes, clouds, or a highly absorptive and scattering atmosphere can prohibit a direct view of a planet's surface across a very broad range of wavelengths. In the solar system, the surface of Venus is obscured because visible and IR radiation is scattered or absorbed and emitted from far above the surface (Titov *et al.*, 2013). Multiple cloud layers of sulfuric acid droplets scatter light, whereas collision-induced absorption and far-wing pressure broadening of $CO_2$ at lower altitudes cause significant IR opacity.

Two key parameters that govern a rocky exoplanet's average climate are the total stellar irradiance $S_*$ (in W/m$^2$) at the exoplanet's orbital position—discussed previously as an "external" property (Table 1)—and the exoplanet's Bond albedo, $A_B$, considered here as an "internal" property critical for the exoplanet's surface temperature (Table 2). These two parameters determine the planet's effective radiating temperature $T_{\text{eff}}$. Then, $T_{\text{eff}}$, with the planet's greenhouse effect, determines the average surface emission temperature $T_{\text{surfe}}$. These quantities are related as follows (see, *e.g.*, Catling and Kasting, 2017, pp 34–36):

$$T_{\text{surfe}} = T_{\text{eff}} + \Delta T_{\text{greenhouse}} = \left[(1-A_B)\frac{S_*}{4\sigma}\right]^{1/4} + \Delta T_{\text{greenhouse}}, \tag{13}$$

where $\sigma$ is the Stefan–Boltzmann constant ($5.67 \times 10^{-8}$ W/(m$^2$·K$^4$)) and $\Delta T_{\text{greenhouse}}$ is the temperature increase of the average surface emission temperature due to the greenhouse effect of the exoplanet's atmosphere. More complicated expressions for $T_{\text{eff}}$ can account for large orbital eccentricities (Mendez and Rivera-Valentin, 2017), but we note that substantial liquid water oceans have large thermal inertia and will likely moderate the effect of eccentricity on habitability even if oceans are periodically ice covered (Kane and Gelino, 2012).

Clearly, the exoplanet effective blackbody temperature $T_{\text{eff}}$ can be calculated if the Bond albedo and incoming stellar flux are known. On the modern Earth, values of parameters in Eq. 13 are $S_* = 1361$ W/m$^2$ (Kopp and Lean, 2011), $A_B = 0.3$ (Palle *et al.*, 2003; Stephens *et al.*, 2015), $T_{\text{eff}} = 255$ K, $T_{\text{surfe}} = 288$ K, and $\Delta T_{\text{greenhouse}} = 33$ K (Schmidt *et al.*, 2010). Although exoplanet surveys frequently assume a default Bond albedo around 0.3 to estimate the effective temperature of a rocky planet (*e.g.*, Sinukoff *et al.*, 2016), in fact, planets with atmospheres in our solar system have a wide range of Bond albedos from 0.76 on Venus (Moroz *et al.*, 1985) to 0.25 on Mars (Pleskot and Miner, 1981). Moreover, the extremes of Bond albedo in the solar system are comets at ∼0.04 and Enceladus at 0.89 (Howett *et al.*, 2010; Encrenaz, 2014). Bond albedo is a more important parameter for Earth-like exoplanets than for giant planets because of surface temperature being key for habitability.

In addition to Bond albedo and stellar flux, knowledge of the concentrations of greenhouse gases, such as those listed in Table 3, is also needed to calculate $\Delta T_{\text{greenhouse}}$ in Eq. 13 and thus surface temperature. Greenhouse gas information would be particularly important in cases wherein direct thermal emission of a planet's surface cannot be measured because the overlying atmosphere is opaque across the measured spectrum. Moreover, if data on $CO_2$, $H_2O$, and surface temperatures can be collected for rocky exoplanets, these could statistically test the HZ hypothesis even before looking for subtle biosignatures (Bean *et al.*, 2017). The calculated extent of the conventional HZ relies upon the idea that the carbonate–silicate thermostat [originally proposed by Walker *et al.* (1981) for the Earth] will apply to other HZ rocky

Table 3. Specific Atmospheric Substances, Their Spectral Bands in the UV-Visible to Thermal-Infrared, and Their Significance for Providing Environment Context for Establishing Whether an Earth-Like Exoplanet Is Truly Habitable

| Substance | Spectral band center or feature, μm | Significance for the planetary environment and habitability |
|---|---|---|
| $CO_2$ | 15, 4.3, 4.8, 2.7, 2.0, 1.6, 1.4, 0.1474, 0.1332, 0.119 | Noncondensable greenhouse gas at $T > 140$ K (*i.e.*, except at the outer edge of a conventional HZ)<br>Well mixed gas, enabling retrievals of atmospheric structure<br>Could be an *antibiosignature* if it coexists with a large amount of $H_2$<br>Could be a substrate for biological C fixation |
| $N_2$ | 0.1–0.15<br>For $N_2$–$N_2$: 4.3, 2.15 | Pressure broadening that enhances the greenhouse effect<br>Possible disequilibrium biogenic gas if detected with $O_2$ and a surface of liquid water |
| $O_2$ | 6.4, 1.57, 1.27, 0.765, 0.690, 0.630, 0.175–0.19<br>For $O_2$–$O_2$: 1.27, 1.06, 0.57, 0.53, 0.477, 0.446 | Possible bulk constituent that enhances greenhouse effect through pressure broadening and weak thermal IR absorption (also a possible biosignature and hence also in Table 4). |
| $O_3$ | >15 (rotation), 14.5, 9.6, 8.9, 7.1, 5.8, 4.7, 3.3, 0.45–0.85, 0.30–0.36, 0.2–0.3 | Possible indicator of $O_2$ from which it derives<br>Greenhouse gas |
| $H_2O$ | Continuum, >20 (pure rotation), 6.2, 2.7, 1.87, 1.38, 1.1, 0.94, 0.82, 0.72, 0.65, 0.57, 0.51, 0.17, 0.12 | Condensable greenhouse gas<br>Abundances near saturation inferred from spectral features may suggest a wet planetary surface or clouds |
| CO | 4.67, 2.34, 1.58, 0.128–0.16 | *Antibiosignature* gas<br>May indicate lack of liquid water |
| $H_2$ | 2.12, NIR continuum, <0.08 continuum | *Antibiosignature* gas if a relatively high abundance coexists with abundant $CO_2$<br>If abundant, pressure broadening that enhances the greenhouse effect<br>Greenhouse effect from pressure-induced absorption with self and other key species (*e.g.*, $CO_2$, $CH_4$) |
| $CH_4$ | 7.7, 6.5, 3.3, 2.20, 1.66, <0.145 continuum | Greenhouse gas<br>In the absence of oxidized species, could indicate a reducing atmosphere. Also a potential biosignature (Table 4) |
| $C_2H_6$ | 12.1, 3.4, 3.37, 3.39, 3.45, <0.16 continuum | Together with $CH_4$, in the absence of oxidized species, could indicate a reducing atmosphere |
| HCN | 14.0, 3.0, <0.18 continuum | In the absence of oxidized species, could indicate a reducing atmosphere |
| $H_2S$ | 7, 3.8, 2.5, 0.2 | Potentially volcanic gas |
| $SO_2$ | 20, 8.8, 7.4, 4, 0.22–0.34 | Potentially volcanic gas |
| $H_2SO_4$ (aerosol) | 11.1, 9.4, 8.4, 3.2[a] | Transient behavior potentially indicates active volcanism<br>May indicate an oxidizing atmosphere<br>Climate effects (cloud condensation nuclei; albedo) |
| Organic haze | Continuum opacity in visible-NIR | Indicates a reducing atmosphere with $CO_2/CH_4 < 0.1$<br>May derive from biogenic or abiotic methane<br>Climate effects (antigreenhouse effect; shortwave absorption) |
| Rayleigh scattering | 0.2–1 | May indicate cloud-free atmosphere and help constrain the main scattering molecule (bulk atmospheric composition) |
| Clouds | UV, visible, NIR, TIR | Climate effects<br>Radiative transfer calculations with scattering (Rayleigh and Mie multiple scattering) may constrain cloud particle sizes and possibly composition |

Also noted is the interpretation of potential biosignature gases.
[a]Exact wavelengths depend on the concentration of $H_2SO_4$ and size distribution of the aerosols.
HZ = habitable zone; IR = infrared; NIR = near-infrared; TIR = thermal infrared.

worlds. In this carbonate–silicate hypothesis, a trend is predicted wherein HZ planets near the inner radius of the HZ should have relatively low $pCO_2$, whereas HZ planets near the outer radius of the HZ should have higher $pCO_2$. These trends could be tested statistically with exoplanet atmosphere data and influence the prior, $P(\text{life} | C)$.

The single most important surface feature that could be observed to enhance our confidence in the habitability and possible inhabitation of an exoplanet would be liquid water. In principle, detection of a glint spot of specular reflection indicates the presence of a large body of liquid water (Sagan *et al.*, 1993; Williams and Gaidos, 2008; Robinson *et al.*, 2010). Glint could be observed as an increasing reflectivity in an exoplanet's crescent phase (Palle *et al.*, 2003; Qiu *et al.*, 2003). However, glint alone just indicates a liquid, rather than water *per se*. For example, hydrocarbon lakes on Titan produce glint (Stephan *et al.*, 2010; Soderblom *et al.*, 2012). So the environmental context given by surface temperature remains crucial even if glint is detected.

A second way to infer an ocean is through the detection of polarized reflected light. Unpolarized light incident on an ocean surface will be reflected with some degree of polarization, which peaks at the Brewster angle. This angle depends on the change in refractive index between the atmosphere and ocean, and, for example, is 53° for a terrestrial air–water interface. Consequently, the fraction of the light that is polarized could be used to infer whether an exoplanet has a liquid water ocean (Stam, 2008; Williams and Gaidos, 2008; Zugger *et al.*, 2010).

Finally, a third way to infer an ocean is from a very attenuated light curve of thermal emission from an Earth-like planet with seasonality, which may indicate a high thermal inertia that would require an ocean (Gaidos and Williams, 2004).

An additional desirable surface feature of an exoplanet to determine would be spectral evidence of a silicate surface indicating a rocky planet, with the caveat that this may be a small fraction of surface area if, like the modern Earth, water and vegetation are ubiquitous. The presence of Si-O on airless rocky exoplanets would produce features in thermal IR emission bands (Hu *et al.*, 2012a), but these would be obscured by a thick atmosphere at temperatures typical of habitability. However, high-resolution reflectance spectroscopy might be able to distinguish broad iron-related absorption bands from narrower atmospheric molecular features. Also, a silicate composition might be suggested from observations of possible nearby planets or moons that lack an atmosphere. Possibly, if silicate reflectance indicates a wholly silicate land surface, this might be interpreted as a bare land surface, suggesting a world that is either uninhabited or lacking plant-like life with only microbes at best, similar to Archean Earth.

"Carbon planets," which are hypothesized to form in C-rich and O-poor protoplanetary disks, would also have solid surfaces but are predicted to be water-poor and uninhabitable to life as we know it (Bond *et al.*, 2010; Carter-Bond *et al.*, 2012; Elser *et al.*, 2012; Moriarty *et al.*, 2014). Planet formation models suggest that high C/O ratio disks could produce planets with iron cores, mantles of SiC or TiC, and diamond or graphite crusts. If such planets exist, reactions with carbon of any water delivered to the planet during accretion could preclude the presence of liquid water, although liquid hydrocarbons on the surface might be possible.

3.2.2. *Atmospheric parameters and properties.* An estimate of an exoplanet's bulk atmospheric composition is highly desirable because it affects all aspects of the atmosphere's behavior: radiation, convection, and dynamics needed for Exo-Earth models to accurately calculate the data likelihoods shown in Figure 1. The bulk atmospheric composition must be known to apply an equation of state. If we use the ideal gas law, $P = \rho \bar{R} T$ (for pressure $P$, temperature $T$, and air density $\rho$), the specific gas constant $\bar{R}$ is set by bulk atmospheric composition using $\bar{R} = R/\bar{M} = k/\bar{m}$, where $\bar{M}$ is the mean molar mass (in kg/mol), $\bar{m}$ is the mean molecular mass ($\bar{M} = N_A \bar{m}$, where $N_A$ is Avogadro's number), and $k$ and $R$ are Boltzmann's constant and the molar ideal gas constant, respectively. Along with gravity, the bulk composition also determines the atmospheric scale height, $H = \bar{R}\bar{T}/g$ for a mean temperature $\bar{T}$. Furthermore, bulk composition fixes the specific heat capacity at constant pressure $c_p$, which sets the dry adiabatic lapse rate $\Gamma_{ad}$ (decline of temperature with altitude in the convective part of a dry troposphere), given by $\Gamma_{ad} = g/c_p$.

Molar mass and heat capacity affect the dynamics of an atmosphere, and thus the climate and associated observables. On a tidally locked body, atmospheres with bigger molar mass tend to have larger day–night temperature differences, narrower super-rotating jets, and smaller zonal wind speeds with more latitudinal variation (Zhang and Showman, 2017). Hence, the bulk atmospheric composition could affect thermal phase curves, and thus likelihoods of photometric data with and without life (Fig. 1).

Atmospheres on Earth-like habitable planets are expected to contain $CO_2$, which is an important gas for a number of reasons. On Earth, environmental feedbacks on $CO_2$ maintain long-term habitability over $\sim 1$ m.y. timescales through the carbonate–silicate geochemical cycle (Walker *et al.*, 1981; Kasting and Catling, 2003; Krissansen-Totton and Catling, 2017; Krissansen-Totton *et al.*, 2018a). But detection of $CO_2$ is also valuable because it should be a well-mixed atmospheric constituent that allows retrieval of atmospheric temperature structure from spectral data, which is necessary to obtain the abundances of other gases, including possible biosignature gases (see section 3.3). In addition, on a habitable world, the presence of $CO_2$ indicates a possible substrate for biological carbon fixation (*e.g.*, Raven *et al.*, 2012).

Other possible bulk gases, such as $N_2$, are more challenging to detect. For $N_2$ partial pressures >0.5 bar, collisional pairs of nitrogen, that is, $N_2$–$N_2$, produce a detectable absorption signal in transit transmission or reflectance spectra at 4.15 μm in the wings of the strong 4.3 μm $CO_2$ absorption band (Schwieterman *et al.*, 2015b) (Table 3).

It is also worth considering the possibility of detecting bulk antibiosignature gases that may be abundant in some types of atmosphere because these would greatly increase the likelihood of $P(D | C, \text{no life})$ in Eq. 7. An *antibiosignature* is any substance, group of substances, or phenomenon that provides evidence against the presence of life. One antibiosignature candidate is carbon monoxide (CO) (Zahnle *et al.*, 2008; Gao *et al.*, 2015), which has absorptions at 4.67 (strong), 2.34 (weak), and 1.58 μm (very weak) (Wang *et al.*, 2016). On Earth, CO is a gas that is readily consumed by microbes in the presence of water, serving both as a metabolic substrate and carbon source (Ragsdale, 2004), and it was probably readily consumed from the primitive atmosphere once life developed (Kasting, 2014). Given that CO is easily oxidized or reduced, it is plausible that microbe-like life elsewhere would remove CO and suppress its atmospheric abundance. The presence of abundant CO would also tend to argue against a liquid water surface because the photochemical sink of CO is the OH radical, which atmospheric chemistry should produce in relative abundance from near-UV photolysis if water vapor evaporates from an ocean.

Another potential antibiosignature is the coexistence of abundant $H_2$ and $CO_2$. These gases are a redox couple for microbial $CH_4$ production, which molecular phylogenetic studies suggest may be one of the most primitive metabolisms on Earth (Weiss *et al.*, 2016; Wolfe and Fournier, 2018). The primitive nature of this metabolism suggests that it may have evolved early on Earth; indeed, geochemical evidence suggests its presence by 3.5 billion years ago (Ueno *et al.*, 2006).

Even if atmospheric bulk composition is incompletely known, it might be possible to infer the redox state of an atmosphere. Redox information is desirable because it can indicate the evolutionary stage of an Earth-like atmosphere, as well as Mars-, Titan-, or Venus-like states, and also the most plausible chemical composition of aerosols. The redox state of the atmosphere can also impact the ocean chemistry, climate, and degree to which different metabolisms are beneficial to biology.

''Reducing'' or ''oxidizing'' are two fundamental chemical categories of atmospheres, and knowing whether an exoplanet atmosphere is one category or the other constrains the overall atmospheric chemistry. Reducing atmospheres are relatively rich in reducing gases, such as the hydrogen-bearing gases $H_2$, $CH_4$, $C_2H_6$, and $NH_3$, and also potentially CO, which is predicted to accumulate in some atmospheres (Zahnle *et al.*, 2008; Kasting, 2014; Sholes *et al.*, 2017). Oxidizing atmospheres generally contain $CO_2$, possibly $O_2$, and only trace levels of reducing gases. Reducing and oxidizing are relative terms and we define them here as relative to a neutral reference atmospheric volatile. Thus, $H_2$ is reducing relative to $H_2O$, whereas $O_2$ is oxidizing relative to $H_2O$; further discussion of redox states can be found in Catling and Kasting (2017, Chap. 8).

The redox state of the atmosphere suggests which hazes are possible and could indicate the evolutionary state of an atmosphere. Reducing atmospheres at low or moderate temperatures tend to contain organic hazes, as observed on Titan and the giant planets of the solar system. Oxidizing atmospheres can contain widespread $H_2SO_4$ aerosols but not Titan-like global organic hazes. The Earth began with a chemically reducing atmosphere that became increasingly oxidized and oxidizing over its history (reviewed by Catling and Kasting, 2017). Looking for the spectral signature of reducing gases, as listed in Table 3, could help identify planets similar to the Archean Earth where $O_2$ is absent but the planet is inhabited. Even if the bulk composition cannot be fully determined, it might be possible to infer the redox state from the relative prevalence of oxidizing or reducing gases.

Surface barometric pressure is an indicator of habitability (as sufficient pressure is required to prevent the rapid sublimation of any surface liquid water), and, when combined with $g$, gives the total atmospheric mass. As with any spectrally inferred surface property, surface pressure can only be measured (or inferred) if the continuum level in a spectrum probes the deepest atmospheric levels. In general, the width of pressure-broadened molecular absorption bands can help to constrain atmospheric pressure, but is degenerate with composition, which also affects broadening (Chamberlain and Hunten, 1987, pp 175–176). With some knowledge of composition, the strength of weak bands can also be used to estimate pressure, for example, in early remote sensing of the solar system, the $CO_2$ content found from the weak 1.6 μm band of $CO_2$ was used to estimate the surface pressure on Mars (Owen and Kuiper, 1964). For reflected-light observations, Rayleigh scattering can indicate pressure, as can the detection of Raman scattering at UV wavelengths (Oklopčić *et al.*, 2016). In transmission spectra, Rayleigh scattering can set a lower limit on atmospheric pressure, given that transmission does not probe down to the planetary surface or below clouds (Lecavalier Etangs *et al.*, 2008; Benneke and Seager, 2012). Also, collision-induced absorption features, where the opacity depends on the number densities of the colliding molecules, have been proposed as barometers due to their pressure dependence (Misra *et al.*, 2014). Finally, for the specific case of tidally locked rocky exoplanets, the day–night temperature gradient depends on surface pressure, equilibrium temperature, and composition (through specific gas constant $\bar{R}$ and heat capacity), so that, in principle, thermal phase curves can be used to infer surface pressure (Koll and Abbot, 2016).

Clouds and hazes are observed on all planets in the solar system with thick atmospheres, although planets such as Earth are much less cloudy than giant planets. Clouds occur when a condensable constituent supersaturates and forms particles (*e.g.*, water vapor turning to liquid droplets when uplifted to a colder altitude), whereas hazes are condensable species produced by photochemical reactions (*e.g.*, organic smog on Titan) (Marley *et al.*, 2013; Horst, 2017). Clouds indicate the presence of a volatile phase, which could be water if spectra are consistent with liquid water drops or water ice particles. Clouds also provide information on the temperature–pressure profile of the planet, as they indicate a liquid or solid phase of a species in the equilibrium state at a layer in the atmosphere. This information could be valuable if the altitude (or pressure level) of the clouds can be determined. A possible sign of clouds, although nonunique, is albedo. Cloudy planets, like those in the solar system, tend to have relatively high geometric albedos (Demory *et al.*, 2013; Marley *et al.*, 2013). Brightness variability at timescales distinct from the rotational timescale could also indicate clouds, whose distributions change due to weather or climate.

The presence of clouds or hazes has consequences for biosignature detection. Clouds and hazes tend to diminish the sensitivity of transit spectra to the deep atmosphere (Knutson *et al.*, 2014; Kreidberg *et al.*, 2014; Robinson *et al.*, 2014; Robinson, 2018). Also, substantial cloud cover can mute absorption features in reflectance spectra (Kaltenegger *et al.*, 2007; Rugheimer *et al.*, 2013).

Hazes could be the photochemical products of volcanic or biogenic gases, which in certain chemical cases could indicate a volcanically active body (Kaltenegger *et al.*, 2010; Misra *et al.*, 2015) or a potentially inhabited early Earth-like body (Arney *et al.*, 2016), respectively. A $1/\lambda^4$ slope (where $\lambda$ is wavelength) of the continuum in the UV-visible part of the spectrum could indicate Rayleigh scattering of particular molecular or aerosol species and possible transparency of an atmosphere (*i.e.*, high single scattering albedo) caused by the haze-free chemistry of an oxidizing atmosphere. For example, such scattering by gases in a relatively haze-free atmosphere causes the Earth to appear from a great distance as a "Pale Blue Dot" (Sagan, 1994; Krissansen-Totton *et al.*, 2016b). Conversely, transmission and scattering characteristics can also indicate the presence of hazes. In transit transmission spectroscopy, a long pathlength means that a haze with only modest vertical optical depth can cause substantial broadband opacity (Fortney, 2005). Thus, the spectrum produced by the scattering of a haze is very different to a purely gaseous medium (Knutson *et al.*, 2014; Kreidberg *et al.*, 2014; Robinson *et al.*, 2014).

"Atmospheric structure" is scientific shorthand for the vertical temperature structure (because pressure and density are then defined by hydrostatics and an equation of state), and the better the thermal structure is known, the better one can retrieve gas abundances from thermal IR spectral data, which may be critical for simulating data for the data likelihoods shown in Figure 1. The structure of strong absorption bands of a well-mixed gas is required to retrieve atmospheric structure. In the NIR, $CO_2$ has a fundamental absorption at 4.3 μm, strong absorptions at 2.7 and 2.0 μm, and weaker absorptions at 5.2, 4.8, 1.6, and 1.4 μm (Table 3). With sufficient spectral resolution, strong $CO_2$ absorption bands can be inverted to retrieve the vertical temperature profile above the surface of a planet or above a cloud surface (*e.g.*, Twomey, 1996; Rodgers, 2000). Once a thermal profile is calculated, one can retrieve the abundances of minor species that could provide information about the environment (the presence of abundant $H_2O$) or potential biosignatures (*e.g.*, $O_3$ or $CH_4$) (*e.g.*, Drossart *et al.*, 1993).

Key trace species in the atmosphere could provide important insights into habitability and influence likelihoods of the data with and without life shown in Figure 1. Earth's reflectance spectrum contains many water vapor absorption features. We should expect a habitable exoplanet with surficial liquid water to have substantial gas-phase $H_2O$ in its atmosphere as a result of evaporation and so numerous $H_2O$ vapor absorption bands should be present. In the NIR, key $H_2O$ absorption bands are centered at 2.7, 1.87, 1.38, 1.1, 0.94, 0.82, and 0.72 μm. Analysis of water vapor's vibration–rotation bands could be used to infer a large $H_2O$ vapor abundance in the atmosphere. If the planet's surface temperature is suitable, a large amount of water vapor could only be explained by evaporation from a large body of liquid water on a planet's surface rather than the cool top of a steam atmosphere.

Liquid or frozen water on a surface has spectral features that are broadened and shifted to different wavelengths compared with water vapor. Stretching vibrations are shifted to longer wavelengths in the condensed phase: from 2.66 and 2.73 μm in water vapor to 2.87 and 3.05 μm in liquid, to 3.08 and 3.18 μm in ice (Fletcher, 1970, his Table 2.1). Formation of a hydrogen bond weakens the "spring constant" for the covalent bond. But hydrogen bonding constrains bending, so the bending "spring constant" becomes stiffer and bending vibrations are shifted to shorter wavelength: from 6.27 μm in vapor to 6.10 μm in ice (Warren and Brandt, 2008). Condensed phases also have much smoother absorption spectra compared with the jagged line spectra of gases (*e.g.*, see Wozniak and Dera, 2007, p 54). Loosely, one can think of this as extreme collision broadening, since in a condensed phase, the molecules are in contact, so always "colliding."

Other trace gases could potentially indicate volcanism, such as $SO_2$, $H_2S$, and HCl (Kaltenegger *et al.*, 2010). The lack of $SO_2$ and HCl spectral absorption has been used as evidence for absence of volcanic activity on Mars (Hartogh *et al.*, 2010; Krasnopolsky, 2012; Khayat *et al.*, 2017). Conversely, such gases could be used to look for active volcanism. In both oxic and weakly reducing atmospheres, some $SO_2$ and $H_2S$ can be oxidized to a sulfate aerosol haze. Consequently, transient increases in NIR extinction in transmission spectroscopy due to aerosol loading could be a signature of pulses of explosive volcanism on an exoplanet (Misra *et al.*, 2015). $H_2SO_4$ itself also has absorption features in the thermal IR and NIR (Palmer and Williams, 1975; Pollack *et al.*, 1978).

### 3.3. Component 3: search for potential biosignatures in the spectral data

Future telescope projects and missions will search for potential biosignatures in the data from habitable worlds, which is the third component of our procedural strategy and provides the basic data that we seek to evaluate in Figure 1. Assessing whether the planet is truly habitable with liquid water on the surface is part of the "internal" climatic, atmospheric, and surface properties of an exoplanet (component 2 shown in Fig. 2), whereas for component 3 we are focused on candidate biogenic constituents including gases, aerosols, or surface pigments that have a measureable remote sensing signature.

The field of remotely detectable biosignatures is still in its infancy, however, and Walker *et al.* (2018, in this issue) consider alternative biosignatures and classes of biosignature. Here, we focus on commonly accepted biosignatures and specific examples. Component 3 in our procedural strategy identifies the spectral signatures of a variety of biogenic molecules, and then determines whether their biological nature is corroborated by the environmental context or the presence of additional biosignatures.

#### 3.3.1. Candidate biogenic gases and their concentrations or column abundance.
Based on the detectability of Earth-based life, it is expected that the most detectable biogenic substances in exoplanetary spectra will be gases. For practical purposes, a *biosignature gas* is one that accumulates in a planet's atmosphere to a detectable level (*e.g.*, Seager *et al.*, 2013; Seager and Bains, 2015). Although living organisms produce many gases, if there is no physical scenario where a certain gas could reach high enough concentrations in an exoplanet atmosphere to be detectable, then it is not an effective biosignature. For example, around a Sun-like star, photochemical destruction prevents the build up of remotely detectable levels of many organic sulfur gases for biospheric fluxes into the atmosphere that are comparable with those of the modern Earth (Domagal-Goldman *et al.*, 2011). Similarly, ammonia has difficulty accumulating to detectable concentrations on planets with Earth-like amounts of radiation from their host stars, even in relatively reducing atmospheres (Sagan and Chyba, 1997).

A variety of proposed spectral gaseous biosignatures are shown in Table 4. Some of these gases also have atmospheric or climatic significance and so were previously listed in

TABLE 4. POTENTIAL BIOSIGNATURE GASES AND ASSOCIATED INFORMATION

| Biosignature | UV-Visible-NIR band center, $\mu m$ and $(cm^{-1})$ | Visible-NIR band interval, $cm^{-1}$ | Thermal IR spectral band center, $\mu m$ | Biogenic source | Abiogenic false positive |
|---|---|---|---|---|---|
| $O_2$ | 1.58 (6329)<br>1.27 (7874)<br>1.06 (9433)<br>**0.76 (13158)**<br>0.69 (14493)<br>0.63 (15873)<br>0.175–0.19 [Schumann–Runge] | 6300–6350<br>7700–8050<br>9350–9400<br>**12850–13200**<br>14300–14600<br>14750–15900 | — | *Photosynthesis*: splitting of water | Cases of water and $CO_2$ photodissociation and preferential escape of hydrogen, with lack of $O_2$ sinks |
| $O_3$ | 4.74 (2110)<br>3.3 (3030)<br>0.45–0.85 [Chappuis]<br>0.30–0.36 [Huggins]<br>0.2–0.3 [Hartley] | 2000–2300<br>3000–3100<br>10600–22600 | >15 (rotation), 14.3, 9.6, 8.9, 7.1, 5.8 | *Photosynthesis*: photochemically derived from $O_2$ | As above |
| $CH_4$ | 3.3 (3030)<br>**2.20 (4420)**<br>1.66 (6005)<br><0.145 continuum | 2500–3200<br>**4000–4600**<br>5850–6100 | 6.5, 7.7 | *Methanogenesis*: reduction of $CO_2$ with $H_2$, often mediated by degradation of organic matter | Geothermal or primordial methane |
| $N_2O$ | 4.5 (2224)<br>4.06 (2463)<br>2.87 (3484)<br>0.15–0.20<br>0.1809, 0.1455, 0.1291 | **2100–2300**<br>2100–2800<br>3300–3500 | 7.78, 8.5, 16.98 | *Denitrification*: reduction of nitrate with organic matter | Chemodenitrification but not truly abiotic on Earth[a]; also strong coronal mass injection affecting an $N_2$–$CO_2$ atmosphere[b] |
| $NH_3$ | 4.3<br>3.0 (3337)<br>2.9 (3444)<br>2.25, 2, 1.5, 0.93, 0.65, 0.55, 0.195, 0.155 | **2800–3150** | 6.1, 10.5 | *Ammonification*: Volatilization of dead or waste organic matter | Nonbiogenic, primordial ammonia |
| $(CH_3)_2S$ | 3.3 (2997)<br>3.4 (2925)<br>0.205, 0.195, 0.145, 0.118 | **2900–3100** | 6.9, 7.5, 9.7 | Plankton | No significant abiotic sources |
| $CH_3Cl$ | 3.3 (3291)<br>3.4 (2937)<br>0.175, 0.160, 0.140, 0.122 | **2900–3100** | 6.9, 9.8, 13.7 | Algae, tropical vegetation | No significant abiotic sources (Keppler *et al.*, 2005) |
| $CH_3SH$ | 3.3 (3015)<br>3.4 (2948)<br>0.204 | **2840–3100** | 6.9, 7.5, 9.3, 14.1 | *Mercaptogenesis*: Methanogenic organisms can create $CH_3SH$ instead of $CH_4$ if given $H_2S$ in place of $H_2$ (Moran *et al.*, 2008). | No significant abiotic sources |
| $C_2H_6$ | 3.37 (2969)<br>3.39 (2954)<br>3.45 (2896)<br><0.16 | **2900–3050** | 6.8, 12.15 | Photochemically derived from $CH_4$, $CH_3SH$, and other biologically produced organic compounds | Could be derived from geothermal or primordial methane |

Shown are absorption band centers or band ranges in the UV-visible to NIR, as well as thermal IR. Particularly strong bands are marked in bold because of their strength and/or lack of contamination from other gases. Square brackets contain the names of particular bands.

[a]$N_2O$ has been generated from "chemodenitrification," whereby nitrite ($NO_2^-$) or nitrate ($NO_3^-$) reacts with $Fe^{2+}$-containing minerals in brines (Jones *et al.*, 2015; Samarkin *et al.*, 2010). However, on Earth, the source of natural oxidized nitrogen ultimately comes from nitrifying bacteria or atmospheric chemistry that relies upon oxygen, which comes from photosynthesis. Also $N_2O$ can be released from UV photoreduction of ammonium nitrate (Rubasinghege *et al.*, 2011), where the latter comes from humans as industrial fertilizer. Another $N_2O$ source comes from very weak *in situ* atmospheric gas phase reactions.

[b]Airapetian *et al.* (2016).

**16**                                                                                                                                                         **CATLING ET AL.**Table 3. The biogenic nature of each species is informed by our knowledge of terrestrial biology. However, in general, a redox reaction as a source of metabolic energy, whether derived from environmental chemical gradients or from excitation of electrons by photons, is a universal aspect of chemistry that should apply to life anywhere. Consequently, photosynthetic $O_2$ (from oxidation of water to obtain electrons and hydrogen to reduce $CO_2$ to organic carbon) or $N_2O$ (from reducing oxides of nitrogen with organic carbon) are ways that microbe-like entities anywhere could use energy for carbon fixation or other biological processes.

Most currently accepted biosignatures consist of spectral evidence of a biogenic gas molecule. For gases, a distinct spectral feature or number of features would identify the presence of a potential biogenic gas and further analysis would deduce a concentration or column abundance of the molecule in an exoplanet atmosphere.

The abundance of a potentially biogenic gas can be determined most simply in the UV, visible or NIR where reflected light dominates over thermal emission from a planetary atmosphere and surface, so that thermal IR can be ignored. For example, ignoring scattering, the one-way zenith optical depth of the $O_2$ A-band (at 760 nm) is related to $O_2$ column abundance $N_0$ (molecules/cm$^2$), given an absorption cross-section $\sigma_{760}$ (cm$^2$/molecule), by

$$\tau_{760} = \int_0^\infty \sigma_{760} n(z) dz \approx \sigma_{760} \int_0^\infty n(z) dz = \sigma_{760} N_0, \quad (14)$$

where number density $n$ ($O_2$ molecules/cm$^3$) is assumed well mixed over altitude $z$. For an exoplanet, if starlight passes through the atmosphere and is reflected at the surface, then the total $O_2$ column in the line of sight, $N_0$, at the center of the disk is less than the amount $N$ seen by an external observer of the entire exoplanet disk according to an "effective airmass" correction, defined as

$$M^* = N/N_0. \quad (15)$$

Owen (1980) calculated that moderate spectral resolution ($\lambda/\Delta\lambda = 20$–100) could resolve the $O_2$ A-band on a planet with a level of $O_2$ the same as the modern Earth, assuming a value of $M^*$ of 3.14. However, the optimum resolution for observing this band will depend on the noise properties of the system. More recent estimates suggest that lower noise instruments can detect this band with spectral resolutions ~150 (Brandt and Spiegel, 2014).

In the general case, an effective airmass will depend on an exoplanet's phase (*e.g.*, full or quarter disk illumination) and the way brightness varies across the disk. Traditionally, one would deduce abundances using the equivalent width, $W$, of a spectral band,

$$W \propto (\sigma_\nu N_0 M^* \Delta\nu_L)^n, \quad (16)$$

where $\Delta\nu_L$ is the line width and $n$ takes values from end members of $n = 1$ (weak lines) to $n \sim 0$ (line saturation just beginning) to $n = \frac{1}{2}$ (strong lines). If unsaturated and saturated lines of a constituent are measured, "curve of growth" laboratory data (of the increasing absorption with mass path of absorber) can be used to find the column abundance of the gas, $N_0$. In principle, if the line width $\Delta\nu_L$ comes from pressure broadening, then, assuming temperature $T$ is known or can be estimated, one can calculate a total pressure $p$ and a volume mixing ratio of a biosignature gas, $c_i = p_i/p$, where $p_i$ is the partial pressure of the absorbing gas, defined by the ideal gas law and Dalton's law:

$$p_i = n_i kT = \frac{N_0}{H} kT, \quad (17)$$

where $H$ is the scale height, which will be the same for the constituent as the bulk atmosphere if it is well mixed.

In a more modern treatment, we know enough about the spectral structure of bands to account for line saturation and overlap explicitly to determine gas abundances from the spectrum of multiple lines. Thus, iterative procedures would fit a spectrum to a line or multiple lines to estimate the gas abundance(s) if the spectral resolution and signal-to-noise are sufficient.

At longer wavelengths, where thermal emission by the surface and atmosphere dominates over reflected light, the shape of strong absorption bands of vertically well-mixed gases (such as $CO_2$ for an Earth-like world) makes it possible to determine the temperature structure and retrieve the abundances of gases from altitude- and temperature-dependent IR emission contributions (weighting functions), as mentioned in section 3.3.2.

Techniques for detecting the gaseous components of an atmosphere and the structure of the atmosphere have been demonstrated on larger, uninhabitable planets (Benneke and Seager, 2012; Barstow *et al.*, 2013; Lee *et al.*, 2013; Line and Yung, 2013; Line *et al.*, 2013; Line *et al.*, 2014; Waldmann *et al.*, 2015; Feng *et al.*, 2016; Rocchetto *et al.*, 2016). Similar retrievals are being considered for the search for biosignatures on Earth-like worlds (von Paris *et al.*, 2013; Barstow and Irwin, 2016; Barstow *et al.*, 2016). In section 3.3.4, we discuss the need for even more sophisticated forms of forward and inverse models that link the atmosphere to the potential biogeochemistry of the planet and associated biosignature metrics.

3.3.2. *Potential surface biosignatures.* In addition to gaseous atmospheric biosignatures, some organisms produce pigments with characteristic spectral reflectance features (Table 5). Such spectral signatures could increase the confidence of the discovery of life on exoplanets. As described by Schwieterman *et al.* (2018, in this issue), on Earth, the vegetation "red edge" is recognized as a globally detectable signature of life, which is a sharp increase in reflectance at ~0.67–0.76 μm (Sagan *et al.*, 1993; Seager *et al.*, 2005). This particular biosignature also exhibits strong temporal variability due to seasonal changes in vegetation distribution and density, adding further observational potential (Schwieterman *et al.*, 2015a).

The vegetation red edge remains the only well-accepted surface biosignature and is caused by red absorbance by chlorophyll *a* in plant leaves and scattering in the NIR where the pigment does not absorb. The red edge is an expression of light harvesting for photosynthesis as well as of the structure of the host organism (see Schwieterman *et al.*, 2018, in this issue). Among phototrophs, the red edge is not unique in expressing these functions but it is widespread over Earth's surface and lacks any strong false positives for life.

Earth-observing satellites quantify the red edge by using the contrast between visible and NIR spectral regions. The most popular metric is the *normalized difference vegetation index* (NDVI) = $(R_{NIR} - R_{RED})/(R_{NIR} + R_{RED})$ (Huete *et al.*, 1994; Myneni *et al.*, 1995; Myneni *et al.*, 1997; Tucker *et al.*, 2005), where $R_{RED}$ is the surface reflectance in the red band and $R_{NIR}$ is the reflectance in the NIR band. The red and NIR bands used for NDVI calculations can vary according to satellite sensor design. Precisely designed sensor bands avoid the range 700–760 nm where the red edge is steep, and target adjacent red and NIR bands to measure the strongest contrast in reflectance. For the red, the band is best centered around the peak pigment absorbance at 650–680 nm. The Earth-observing satellites Moderate Resolution Imaging Spectrometer (MODIS) and the Landsat Thematic Mapper (TM), which were designed to map vegetation, use a 630–690 nm red band. The index may also sometimes be calculated by using a broadband visible (400–700 nm) absorbance in place of the red band, since plants harvest light across the visible. For the NIR band, plant leaves have an NIR reflectance plateau across ~760–1100 nm; however, this range is also impacted by water absorbance bands. MODIS and Landsat TM precisely target a 780–900 nm NIR band to avoid overlap with the red edge and with water.

In an exoplanet observation, a red edge signal will not necessarily be atmospherically corrected in telescopic observations of a disk average spectrum. Arnold (2008) investigated the red edge in the Earthshine defining an index VRE = $(R_{NIR} - R_{RED})/R_{RED}$, which provided a simple measure of the NIR reflectance bump in the absence of atmospheric correction (VRE is not a standard index in vegetation remote sensing and Arnold used the reflectance in a narrow NIR spectral range [740–800 nm] for $R_{NIR}$ and reflectance in a narrow red band [600–670 nm] for $R_{RED}$). They found that the red edge detectability ranged from 1% to 12% in observations and models. For space-based telescopes such as the Terrestrial Planet Finder/Darwin concepts (Beichman *et al.*, 2007; Defrère *et al.*, 2017), the exposure times would be ~100 h to reach a sufficient spectral precision (<3%) with a spectral resolution ($\lambda/\Delta\lambda$) of 25 for an Earth at 10 parsecs distance (Arnold *et al.*, 2002). If detected, seasonal variations would be especially informative.

Besides the red edge, life produces a great diversity of pigments from phototrophic light-driven transmembrane proton translocation to photoreactive, phototactic, photoprotective, and photorepair systems, respiration components, and photosynthetic reaction centers (Schwieterman *et al.*, 2015a). The absorption maxima varies from UV to NIR (0.1–2.5 μm). Pigments that absorb in the visible include, for example, cytochromes (~0.40 μm), carotenoids (0.28–0.55 μm), cyanobacterial and chloroplast accessory pigments (0.40–0.74 μm), bacteriochlorophyll pigments (0.74–1.03 μm), and retinal pigments (0.53–0.63 μm) (DasSarma, 2006; Hegde *et al.*, 2015). Moreover, chlorophyll pigments have recently been discovered in cyanobacteria, tuned to absorb in the far red/NIR as long ~0.77 μm (Chen, 2014; Li and Chen, 2015). Biogenic surface spectral features are also the result of ecological signaling (Schwieterman *et al.*, 2015a), community signatures (Parenteau *et al.*, 2015), chiral molecules (Sparks *et al.*, 2009; Sparks *et al.*, 2012), and the bidirectional reflectance distribution function that accounts for angle-dependent reflectivity (Doughty and Wolf, 2010). On Earth, many of these biological surface features are only detectable in local remote sensing data, and whether any of these features could dominate on a global scale on an exoplanet remains an open question.

Essentially, multiple issues remain to be addressed for red edge and pigment biosignatures. (1) Could red edge-like biosignatures appear in another wavelength range because of adaptation to a different stellar spectrum or is the wavelength of the red edge feature constrained by molecular limits (Kiang *et al.*, 2007a, 2007b)? For example, could oxygenic photosynthesis be excluded on exoplanets orbiting very cool M dwarf stars because the star's output is mainly in the NIR, which excites vibrational transitions rather than the electronic transitions needed for photosynthesis? (2) If such features exist, what are their false positives? Could a feature similar to the red edge at a different wavelength have false positives from mineral features? There is not yet a consensus on why the vegetation red edge is red, although some argue that chlorophyll *a* evolved to exploit the large number of photons in the red part of the Sun's spectrum convolved with the free energy that each converted photon contributes and transmission of Earth's atmosphere (Bjorn, 1976; Milo, 2009; Marosvolgyi and van Gorkom, 2010; Björn and Ghiradella, 2015). (3) If a different pigment signature emerged on another planet, would it also be unique? Clearly, pigments whose role is light harvesting and screening of ionizing radiation or excess light must have a spectral absorbance that depends on the available light. However, the prominent wavelength bands will rely on molecular energy transduction constraints and the efficiencies, which result from contingent evolutionary pathways, including competitive survival strategies. Also, some pigments adapt colors in real time to fluctuating colors of light (Kehoe and Gutu, 2006). The colors of other pigments may be fortuitous without a clear function with respect to light interactions. Often, many of these pigments, such as carotenoids in halophilic Archaea, may be expressed in response to environmental chemistry, such as oxygen concentration or pH, in addition to light intensities (Slonczewski *et al.*, 2010). Moreover, some pigments may be stratified and tuned depending on the environmental conditions.

A quantitative framework for evaluating surface biosignatures remains unclear. The possible adaptations of a red edge-like signature and evaluating other pigments and surface biological features as potential biosignatures serve as motivating cases for developing a framework for identifying biosignatures that are functions of their environmental context (Walker *et al.* 2018, in this issue). Except for chirality and the terrestrial vegetation red edge, the uniqueness of the other surface spectral signatures and potential for false positives have yet to be scrutinized.

**3.3.3. Given the environmental context, are detected species conceivably biogenic?** We have already discussed gaseous and surface biosignatures, but what corroborating observations can strengthen or weaken the possibility that life has been found? We expand on two types of corroboration: if the planetary environment and atmospheric environment are conducive to life, and/or if there is direct, supporting evidence of biology itself. Environmental information could be used to improve estimates of data likelihoods or alter the priors shown in Figure 1, whereas new biosignature data would add to the ''data'' box of Figure 1.

*3.3.3.1. Does the planetary environment and atmospheric environment support a biological source?* The most fundamental evidence supporting inhabitation would be a surface emission temperature, $T_{surfe}$, that allows stable liquid water on an exoplanet, which would increase the prior $P(\text{life} \mid C)$ and improve modeling of the data likelihoods shown in Figure 1. Alternatively (or additionally), the inference of surficial liquid water could come from abundant $H_2O$ vapor lines in the spectrum, evidence of glint, or polarized reflected light (discussed in section 3.3.2). Color variations on a rotating planet might provide circumstantial support for life and would be additional photometric ''data'' shown in Figure 1.

Further environmental corroboration would be the presence of potential substrates or side products for metabolisms that are being proposed as responsible for a biosignature gas. On an oxidized world, photosynthesis requires $CO_2$ as a chemical substrate, so detection of $CO_2$ would support the possibility that $O_2$ is a biosignature. Biological denitrification produces $N_2O$, but (on Earth, at least) it also produces even more $N_2$, so that the presence of the two gases together might support the idea that $N_2O$ is biological.

The presence of a biosignature gas should also be accompanied by its expected photochemical products. For example, the detection of $O_2$ would be corroborated by the detection of $O_3$ in the atmosphere because $O_3$ is a photochemical product of $O_2$, if the star produces adequate near-UV flux. Such pairs of detections can also be used to help constrain gas concentrations—for example, the $O_3$ amount depends on the $O_2$ concentration, although nonlinearly (Kasting *et al.*, 1985).

The mixture of gases in an atmosphere could be consistent or inconsistent with biology, and with the biosignature that has been detected. If a biosphere modulates the composition of a planetary atmosphere, then it will tend to remove gases that are easy to metabolize. In a reducing atmosphere analogous to that of the Archean Earth on an inhabited planet, we would not expect $H_2$ to attain a high abundance in the presence of $CO_2$ because these two substances can be readily used as metabolic reactants to produce $CH_4$ through the methanogenesis pathway, $CO_2 + 4H_2 \rightarrow CH_4 + 2H_2O$. Laboratory experiments and bioenergetic calculations show that >90% of the hydrogen is converted to methane if $CO_2$ is not limiting (Kral *et al.*, 1998; Kasting *et al.*, 2001). In addition, if $CH_4$:$CO_2$ ratios exceed $\sim 0.1$, and there is adequate UV flux from the star, an organic haze is produced (Trainer *et al.*, 2006; Haqq-Misra *et al.*, 2008; Arney *et al.*, 2016), which may decrease biological productivity through an antigreenhouse effect on the climate (Haqq-Misra *et al.*, 2008).

Thus, if the concentration ratio of $CO_2$:$CH_4$:$H_2$ consists of sharply declining numbers, it would be consistent with the presence of biosphere in which methanogenesis is converting $CO_2$ and $H_2$ into $CH_4$. On the current Earth, the ratio $CO_2$:$CH_4$:$H_2$ is $\sim 400:1.8:0.5$, where the numbers are in ppmv. In contrast, a small amount of $CH_4$ in a hydrogen-rich atmosphere (a high $H_2$:$CH_4$ ratio) could result from abiotic chemistry, as found today on the giant planets of the solar system (Irwin, 2009), and the interpretation of $CH_4$ as a biosignature would be doubtful because such atmospheres could be primordial. Similar arguments about $CH_4$:$H_2$ ratios have also been applied to life within the solar system, on both Enceladus (Waite *et al.*, 2017) and Mars (Sholes *et al.*, 2017).

Finally, if there is enough information about the atmospheric composition and planetary environment, it may be possible to quantify atmospheric chemical disequilibrium. On Earth, large biogenic fluxes of gases from different ecosystems produce mixtures of gases that are chemically unstable on long timescales (Lovelock, 1965; Lovelock, 1975). A methodology to quantify thermodynamic chemical disequilibrium has been established and could be applied (Krissansen-Totton *et al.*, 2016a, 2018b). To calculate chemical disequilibrium properly for an Exo-Earth, one must consider all fluid phases because the largest source of disequilibrium in Earth's atmosphere–ocean system is attributable to the coexistence of $N_2$, $O_2$, and liquid water, as quantified by Krissansen-Totton *et al.* (2016a) and qualitatively noted by, for example, Lewis and Randall (1923, pp 567–568) and Emerson and Hedges (2008, pp 96–99). This disequilibrium is biogenic because current levels of both atmospheric $N_2$ and $O_2$ would not persist in the absence of biogenic fluxes of both gases (*e.g.*, Som *et al.*, 2016; Stüeken *et al.*, 2016). For an exoplanet, detecting such multiphase disequilibria would require knowledge that an extensive liquid ocean exists by using techniques such as those described in section 3.3.2.

If just gaseous components are known for an exoplanet, it may be possible to estimate their kinetic instability, that is, residence time against photochemical destruction, and deduce whether a large, potentially biogenic source is needed. A short lifetime ($\sim 10$ years) and an inferred large flux into the atmosphere are the case for $CH_4$ in Earth's current $O_2$-rich atmosphere.

For biological pigments, the planet's environment may help predict the presence of a pigment due to its function, although there are many uncertainties. For example, light-harvesting pigments should absorb at wavelengths of available light with sufficient energy per photon to perform the necessary tasks of charge separation or proton pumping (Kiang *et al.*, 2007b; Stomp *et al.*, 2007). Bacteria such as the radiotolerant *Deinococcus radiodurans* manufacture antioxidant carotenoids, such as deinoxanthin, in response to UV, and extraterrestrial microbe-like life may have similar physiological response (Tian and Hua, 2010). However, some biological pigments have colors that are fortuitous and unrelated to the light environment, but may be a function of the chemical environment, such as pH (Slonczewski *et al.*, 2010). Other pigments may be related to nano- or microstructure and be purely evolutionary contingent. Indeed, some take the view that all biological pigments that are not directly concerned with collecting light energy, for sensing or metabolism, should be considered as arbitrary and unpredictable. Since organisms can have more than one strategy for survival, and we cannot measure environmental variables like pH remotely, constraining the suitable environments for these alternative biosignatures is a matter of more research.

*3.3.3.2. Are there corroborating biosignatures?* If a biosignature is detected, direct corroborating evidence could take the form of either evidence of the biogenic source of that biosignature or there could be another, independent biosignature, suggesting the presence of biology that is not directly related to the other biosignature. We give some examples of these in what follows.

*3.3.3.2.1. Corroborating evidence of a biogenic source.* If $H_2O$ and $O_2$ have been detected, an additional corroborating biosignature would be the red edge of a pigment that is widespread on the planet and essential to the function of photosynthetic production of $O_2$. Also, the presence of biogenic $N_2O$ would be consistent with oxygenic photosynthesis because it derives from abundant organic matter produced from photosynthesis and oxidized forms of nitrogen. The latter include nitrate ($NO_3^-$) or nitrite ($NO_2^-$) that form in the presence of $O_2$ when ammonium ($NH_4^+$) derived from organic matter is oxidized.

Another possibility might be seasonal trends consistent with biology in the photometric data of Figure 1. For example, seasonal variations of a potentially biogenic gas ($CH_4$) and substrate for photosynthesis ($CO_2$), or pigment could be corroborative. Such temporal variations occur on Earth (*e.g.*, Schlesinger and Bernhardt, 2013, Chap. 11).

*3.3.3.2.2. Another, not necessarily directly corroborative, biosignature.* If $O_2$ has been detected, the presence of $CH_4$ would be an example of an independent biosignature, given that $CH_4$ is not expected to be abundant in an $O_2$ atmosphere unless there is a significant source, such as methanogenesis. The simultaneous detection of $O_2$ and $CH_4$ would constitute a very compelling biosignature because of the short photochemical lifetime of $CH_4$ in oxidizing atmospheres, as already described.

Although we have excluded SETI because of a focus on a framework for gas/pigment biosignatures (as mentioned in the Introduction), new HZ exoplanets are potentially subject to SETI observations, so we briefly mention how SETI null or positive observations fit within our framework. A SETI null signature only excludes extraterrestrial intelligence (ETI) and so provides limited prior information in a scheme for detecting biosignature gases from a global biosphere of non-ETI life. For example, a null SETI signal detected at 100 Ma from Earth by an ETI civilization observing our planet remotely should not have affected their priors for the existence of non-ETI life on Earth at all because a null SETI signal excludes ETI only.

A positive SETI signal provides certainty about the presence of ETI, by definition, but does not necessarily indicate that a biosphere or that biosignature gases are present. The latter is because ETI machines can exist on an airless moon or planet. In fact, some authors speculate that artificial intelligence (AI) should reach a runaway point where AI designs better and better future AI so that machines are the logical descendants of a technological civilization. If so, SETI signals may be more likely to come from machine life rather than organic-based ETI (Shostak, 2015). Hence, exoplanet priors for ETI are potentially completely different to the habitable requirements for a non-ETI biosphere that produces gas or pigment biosignatures; ETI priors are also subject to considerably more speculation, given unknown evolutionary and sociological factors.

*3.3.4. The need for chemical-radiative forward models and retrievals of atmospheric composition.* The aforementioned discussion of non-ETI biosignatures indicates the obvious need to understand the biogeochemistry of exoplanets to interpret biosignatures through calculations of data likelihoods and interpret the context for a Bayesian prior, $P(\text{life} \mid C)$, needed for the overall biosignature framework (Fig. 1). Of course, considerable work still needs to be done to understand the coevolution of biogeochemical cycles and atmospheric composition on Earth, which is being addressed by using "Earth System" models. Such models are corroborated through geochemical and isotopic paleoclimate proxies over geologic time (Berner, 2004; Mackenzie and Lerman, 2006; Kump *et al.*, 2010; Langmuir and Broecker, 2012).

For exoplanets, we conceive of "Exo-Earth System" models, as mentioned earlier, that would couple a surface biosphere to atmospheric chemistry and climate. The latter would necessarily use radiative transfer; models could also include atmospheric and ocean dynamics. A model could incorporate how the interior thermal evolution of a planet couples to mantle processes and outgassing of volcanic and metamorphic gases. Ultimately, such forward models would be used to generate a synthetic reflectance or transmission spectrum and Bayesian likelihoods, $P(D \mid C, \text{life})$ and $P(D \mid C, \text{no life})$ (Fig. 1). Hitherto, such spectra have been generated for more specific cases that use Earth-like exoplanet atmosphere models rather than Exo-Earth System models *per se* (*e.g.*, Robinson *et al.*, 2011; Arney *et al.*, 2016).

In the future, an Exo-Earth System model could be configured in the form of an inverse fit to spectra or photometry data from exoplanets (Fig. 1). With data of sufficient fidelity, a Bayesian fit would allow for quantitative constraints on possible gas fluxes that are needed to support a potential biosignature gas. An investigation of covariance of uncertain parameters in such a fit would show which parameters we need to know better from future observations to improve confidence that the spectrum really represents biology. For example, a better constraint of one particular parameter could remove degeneracies (*i.e.*, where two or more parameters have similar overall effects) and produce a better model fit to the spectrum. Such parameters could be particular environmental constraints or atmospheric gas abundances, for example. Further discussion of general developments needed in the area of biosignatures is given by Walker *et al.* (2018, in this issue).

### 3.4. Component 4: Excluding explicit false positives

Situations exist where a molecule that is predominantly biogenic on Earth could be an abiotic species, as noted in Table 4. False positives need to be evaluated, as noted in Figure 2, and quantified by a likelihood $P(D \mid C, \text{no life})$ in our framework (Fig. 1). For example, as reviewed by Meadows *et al.* (2018, in this issue), several theoretical studies have suggested that abiotic, bulk $O_2$ could be present under certain hypothesized circumstances. In one such case, oceans are hypothesized to vaporize and be lost from HZ planets in the luminous premain sequence phases of M dwarfs, leaving behind very $O_2$-rich atmospheres (Luger and Barnes, 2015). There are some skeptics of such ideas, however. Zahnle and Catling (2017) note the tendency of oxygen to photochemically buffer H:O loss at 2:1 ratio at high escape rates. It is notable that neutral $O_2$ is undetectable on Venus, the one case of a postrunaway greenhouse in our solar system, where oxygen must have been either lost efficiently (Zahnle and Kasting, 1986; Zahnle *et al.*, 1988) or taken up by a magma ocean (Hamano *et al.*, 2013), which may also occur on exoplanets (Schaefer *et al.*, 2016). The extent to which purely hypothetical ideas about $O_2$ buildup are realized remains to be determined, but their possibility demands caution. Meadows *et al.* (2018, in this issue) describe the potential observational diagnosis of various hypotheses of abiotic $O_2$.

A general procedue for excluding false positives of biogenic processes might include the following steps. First, it is necessary to identify a plausible abiotic source. Second, the abiotic source must be detected or supported by corroborating evidence. We expand on these concepts, as follows.

3.4.1. *Are there abiotic sources and scenarios where the abiotic source could dominate?* Considerable research has been devoted to identifying abiotic sources of $O_2$ and its photochemical product $O_3$ (Hu *et al.*, 2012b; Tian *et al.*, 2014; Domagal-Goldman *et al.*, 2014; Wordsworth and Pierrehumbert, 2014; Gao *et al.*, 2015; Harman *et al.* 2015). However, relatively little attention has been given to abiotic scenarios where other candidate biosignature gases could be significant. In general, the following processes could cause abiotic sources of gases.

*3.4.1.1. Photochemistry and atmospheric-loss processes (including escape-enhanced gases).* For example, in a dry $CO_2$-rich atmosphere, $O_2$ could potentially build up from photochemical destruction of $CO_2$ (Gao *et al.*, 2015). It has also been proposed that $O_2$ could build up abiotically on highly irradiated planets as a result of escape of hydrogen from water photolysis (Luger and Barnes, 2015).

*3.4.1.2 Geothermal (volcanic, metamorphic, and geomorphological) processes.* Methane is an example of a biogenic gas that may have abiotic sources. More than 90% of $CH_4$ on Earth has a direct biogenic source or comes indirectly from biology through thermal breakdown of old organic matter, but some argue that $CH_4$ can be produced abiotically. The geological process of serpentinization produces $H_2$ when liquid $H_2O$ oxidizes iron in rocks. Fischer–Tropsch reactions could cause subsequent abiotic reduction of inorganic carbon into methane and other abiotic organic compounds (Berndt *et al.*, 1996; Etiope and Lollar, 2013; Guzman-Marmolejo *et al.*, 2013). However, some recent laboratory simulations of serpentinization have produced no detectable $CH_4$ (McCollom and Donaldson, 2016; Grozeva *et al.*, 2017), suggesting that $CH_4$ may have arisen in earlier experiments from organic contamination (McCollom, 2016). Thus, how much $CH_4$ can be produced abiotically on an exoplanet as a false positive remains a matter of research.

*3.4.1.3. Impact delivery.* Impacts that have reducing chemistry in the impact plume could generate substantial amounts of abiotic $CH_4$ and $NH_3$ (Zahnle *et al.*, 2010), which are potential false positives.

*3.4.1.4. Surface signatures: Mineral spectra or other spectral contaminants.* Cautionary tales of spectral contaminants are the false spectral detection of vegetation and methane on Mars in the 1950s and 1960s. Spectral features near 3.5 μm (which became known as ''Sinton bands'') were detected by ground-based telescopes and erroneously attributed to vegetation due to similarity with lichen spectra (Sinton, 1957; Sinton, 1959). Subsequent work showed that the Sinton bands were due to terrestrial water with deuterium, HDO (Rea *et al.*, 1965). Similarly, initial reports of ammonia and methane in the NIR reflection spectrum of the polar caps of Mars observed by the Mariner 7 flyby were subsequently shown to be due to solid $CO_2$ at the surface (Herr and Pimentel, 1969).

*3.4.1.5. Spectral contamination by a moon (or a parent planet if the target body is a moon).* When an exoplanet hosts a moon with its own atmosphere, a single spectrum that includes both moon and exoplanet could end up resembling a mixture of both atmospheres and perhaps could generate chemical disequilibrium false positives (Rein *et al.*, 2014).

*3.4.1.6. Surface sources from mass loss.* Potentially, one could conceive of exoplanets that lose mass from their surfaces in a way that resembles the input of biogenic gas. For example, clathrates in cold regions could melt and input methane into an atmosphere (Levi *et al.*, 2014). Another example is unexpectedly abundant molecular oxygen that was found in the coma of comet 67P/Churyumov-Gerasimenko. Here, $O_2$ may be formed abiotically by collision of energetic water ions with oxygen-containing minerals on the comet's surface (Yao and Giapis, 2017).

3.4.2. *Find corroborating evidence to support an abiotic source.* In general, a false positive tends to occur under circumstances that are at particular extremes of the parameter space we currently expect for rocky, atmosphere-bearing exoplanets and so should be accompanied by diagnostic photometric features associated with those hypothesized conditions. In the proposed cases of abiotic oxygen, various diagnostics are possible (Schwieterman *et al.*, 2016; Meadows, 2017) (see also Meadows *et al.*, 2018, in this issue). If $O_2$ builds up on a dry planet from $CO_2$ photolysis (Gao *et al.*, 2015), the dryness and presence of CO are diagnostics. A proposed build up of tens or hundreds of bar of $pO_2$ surface pressure on planets around M-stars because of a runaway greenhouse during premain sequence superluminous phase (Luger and Barnes, 2015), if real (Zahnle and Catling, 2017), would be indicated by a strong $O_4$ dimer feature (Misra *et al.*, 2014). Hypothetical planets that lack noncondensable $N_2$ (if that is cosmochemically plausible) and build up $O_2$ from water vapor destruction and loss (Wordsworth and Pierrehumbert, 2014) would lack the $N_4$ dimer signatures and may have strong $O_4$. Similarly, in cases of abiotic $O_2$ or $O_3$ photochemical buildup on M-star planets (Domagal-Goldman *et al.*, 2014; Tian *et al.*, 2014), if real (Harman *et al.*, 2015), a lack of $CH_4$ would be expected. Similar diagnostics for other potential false positives are expected and remain to be investigated.

After gathering information pertinent to all four framework components shown in Figure 2, it is likely that follow-up observations would be required, that is, gaining further observations relevant to components 1–4 as necessary. The exact procedure would be defined by the specific scenario and suite of observations existing at the time, which is a matter for future observers, so we do not comment further.

## 4. A Proposed System of Confidence Levels of Life Detection on Exoplanets

The proposed biosignature assessment framework (Figs. 1 and 2) ultimately should produce posterior probabilities for the presence of life on an exoplanet, which can be tied to a system of qualitative descriptions for the confidence of life detection. Similar ideas have been used in other scientific areas. For example, global warming is described in such terms in reports of the Intergovernmental Panel on Climate Change (IPCC). Their set of five basic probability levels are very likely 90–100%, likely 66–100%, about as likely as not 33–66%, unlikely 0–33%, and very unlikely 0–10% (Stocker *et al.*, 2013, their Box TS.1).

Table 5. Surface Biosignatures and Related Information

| Biosignature | Visible-NIR band center, μm | Visible-NIR band interval, cm$^{-1}$ | Biogenic source | Abiogenic false positive |
|---|---|---|---|---|
| Terrestrial vegetation red edge | 0.67–0.76 (a sharp slope between visible absorbance and NIR scattering) | 14925–13160 | *Photosynthesis*: "red edge" due to sharp lack of absorption in the NIR by chlorophyll *a* | None on Earth |
| Carotenoids, retinal and other biological pigments | 0.40–0.50, 0.53–0.63, and absorption over other narrow wavelength ranges | Aromatic C-H 3000–3000<br><br>Schiff base compounds (*i.e.*, $R_2C=NR'$ [$R' \neq H$]) 1620–1700 and others | Photoprotective, photo-trophic, or photoreactive and accessory pigments | Future research |
| Chirality or circular polarization | Dependent on pigment | — | Biological cell molecular structure | Future research |

If high accuracy is known at the upper and lower bounds of probabilities, the IPCC also recommends "virtually certain" for 99–100%, "extremely likely" for 95–100%, "extremely unlikely" for 0–5% probability, and "exceptionally unlikely" for 0–1%. In cases of certainty (*e.g.*, in our case, imagine that an SETI broadcast of a prime number series from an exoplanet corroborates and trumps earlier detection of atmospheric biosignatures on the planet), the confidence level becomes "unequivocal."

We follow the IPCC example with five basic levels of confidence in reporting the possible presence of life on exoplanets (Table 6). A *Level 1* detection of a *very likely inhabited* exoplanet requires multiple lines of evidence and a posterior probability of life being present of 90–100%, given the data and context, that is, $P(\text{life} | D, C) \geq 0.9$ shown in

Table 6. Possible Categories for Probability of Life Detection on Individual Exoplanets

| Confidence level for detection of life | Posterior probability P(life\|data, context) | Evidence | Suggesitve but purely illustrative examples |
|---|---|---|---|
| Level 1: very likely inhabited | 90–100% | Multiple lines of evidence for life. Given current understanding of planetary processes, no known abiotic process can plausibly explain all observed features. | An $O_2$-rich atmosphere with other biosignature gases, including $CH_4$ and $N_2O$, and a liquid ocean identified on an Earth-size exoplanet in the HZ. |
| Level 2: likely inhabited | 66–100% | The body of evidence is consistent with the presence of life. | Atmospheric $O_2$ detected together with $CO_2$ and water vapor on an exoplanet in the HZ. |
| Level 3: about as likely as not inhabited (inconclusive) | 33–66% | Some evidence for life, but insufficient contextual information to draw a definitive conclusion because plausible alternative abiotic explanations cannot be ruled out. | $O_2$ detection in isolation; or an organic haze with abundant $CH_4$; or pigment-like biosignatures; or $N_2$–$CO_2$ atmosphere. Circumstantial evidence for liquid water on a planet in the conventional HZ. |
| Level 4: likely uninhabited | 0–33% | Observational evidence that the planet is habitable, but no biosignatures detected despite an exhaustive search. | Planet is in the HZ and has an atmosphere with abundant water vapor features. But no biosignatures are detected despite extensive data. |
| Level 5: very likely uninhabited | 0–10% | Criteria for habitability are not met or atmospheric antibiosignatures are detected. | $CO_2$-rich, desiccated planets; or $CO_2$–$H_2$ antibiosignature atmosphere; or abundant CO antibiosignature |

Examples are illustrative and based on current thinking in the field. As noted in the second column, specific exoplanet cases would have to be quantified by calculating a Bayesian posterior probability to determine the exact level.

Figure 1. For example, a HZ planet very similar to Earth with coexisting atmospheric $O_2$, $CH_4$, $N_2O$, and $CO_2$, along with evidence of liquid water ocean, might fall into this category because an abiotic system would be very unlikely to mimic all of those data. A *likely inhabited Level 2* detection (at least 66%, or 66–100% posterior probability) would have consistency with the presence of life. For example, an $O_2$-rich atmosphere with a pigment-like spectrum—with no further information—would be suggestive of life, but not as definitive as Level 1, and might qualify for Level 2. A *Level 3* detection (33–66% posterior probability) would be inconclusive, and essentially equivalent to a weaker statement of ''may or may not be inhabited.'' A detection of $O_2$ in isolation might only be a Level 3 detection in this assessment, where some possible alternative abiotic scenarios cannot be ruled out.

If $P(\text{life}|D, C)$ is a maximum of 0.33, an exoplanet should not be reported as inhabited. A *Level 4* planet has *unlikely* inhabitation (0–33% posterior probability) and may require a lack of biosignature detection even though some data allow the planet to be habitable in principle, for example, a planet in the HZ with an atmosphere but no detection of biosignatures despite an extensive search. Finally, a *Level 5* planet (0–10% posterior probability) would have direct evidence that probably rules out its habitability and renders it *very unlikely* inhabited. For example, the exoplanet might be dry with an atmosphere dominated by $CO_2$. Alternatively, the planet might have the presence of antibiosignatures. However, Level 5 ''dead'' planets would be extremely important for understanding the prevalence of life by building up statistics.

## 5. Conclusions

We have presented a general scheme for observing potential exoplanet biosignatures and determining confidence levels for positive detection or nondetection of signs of life (Fig. 1). The proposed framework uses models of planetary atmospheres and environments with spectral or photometric data that contain potential biosignatures to find the Bayesian likelihoods of those data occurring if the exoplanet has, or does not have, life. The latter includes ''false positives'' where abiotic sources mimic biosignatures. Prior knowledge (including all factors that influence habitability and previous exoplanet observations) gives a prior probability of life being present on a potentially habitable exoplanet. The likelihood of data occurring in the presence or absence of life is weighted by that prior probability. The result is a Bayesian posterior probability of life existing on a given exoplanet given the observations and context.

Within this framework, we outlined four observational and analytical components to determine whether biosignatures truly represent the presence of life, which are intended to be iterative, that is, knowledge from some components can be improved or expanded with subsequent observations and models if biosignatures are suspected. The procedural components (shown in Fig. 2) are as follows:

(1) Characterize the stellar properties and exoplanetary system properties to determine whether the planet is in a stable HZ, which informs the prior, $P(\text{life}|C)$, given context $C$. Stellar parameters, for example, age and spectrum, provide further context for interpreting potential biosignatures through evaluating data likelihoods (Fig. 1). The orbital parameters of the exoplanetary system and ''external'' parameters of the exoplanet itself (*e.g.*, mass and radius) are required. Properties that are desirable or essential for this component of the framework are tabulated in Table 1.

(2) Characterize the exoplanet surface and atmosphere to determine whether the surface environment is potentially habitable. These include the thermal emittance and Bond albedo, which are needed to estimate the surface temperature (*i.e.*, whether liquid water is stable on the surface) as opposed to merely being in a HZ. Such ''internal'' properties of an exoplanet that are desirable or essential for this component are tabulated in Table 2, whereas spectral bands for atmospheric composition are tabulated in Table 3.

(3) Search for potential biosignatures in the available data. Most biosignatures consist of spectral evidence of a biogenic gas (Table 4), and a deduction of its concentration or column abundance in an exoplanet atmosphere but time-variable photometry may also be useful (Fig. 1). Pigments and surface biosignatures (Table 5) represent potentially confirmatory biosignatures. The biogenic nature of such molecules must be assessed through the environmental context deduced through components 1 and 2 and also any corroborating evidence such as other information on atmospheric composition. Models of atmospheric composition and climate are an important part of such an assessment. Covariances of parameters identified in inverse modeling with a chemical-radiative atmospheric forward model would help prioritize which parameters are most important for follow-up observations.

(4) Exclude false positives. Coupled geochemical–climate ''Exo-Earth System'' models would be used to simulate the interconnected parts of an Earth-like planet, such as the interior (*e.g.*, mantle thermal evolution), ocean, biosphere, and atmosphere. Critically, Exo-Earth models must be developed that generate synthetic exoplanet spectra or photometric data $D$ to estimate the likelihoods of the data occurring in either the presence or absence of life, $P(D|C, \text{life})$ and $P(D|C, \text{no life})$, given the exoplanet's context, $C$.

The result of the mentioned procedures and Bayesian assessment framework (Fig. 1) is to provide data $D$ and context $C$ for calculating Bayesian posterior probabilities of life detection on an exoplanet, $P(\text{life}|D, C)$. We suggest that such posterior probabilities map to five confidence levels for announcing the results of searches for life on exoplanets (Table 6). With decreasing confidence, these levels and posterior probabilities are ''very likely'' (90–100%), ''likely'' (66–100%), ''inconclusive'' (33–66%), ''unlikely'' (0–33%), and ''very unlikely'' (0–10%) inhabited. We speculated about possible criteria that might determine whether an analysis of future data would fall into a particular level, noting that in reality, detailed calculations would be required to quantify the Bayesian posterior probabilities for each exoplanet data set and context.

## Acknowledgments

D.C.C., D.C., N.Y.K., and S.D.-G. were supported for this work by NASA Astrobiology Institute's Virtual Planetary

Laboratory under Cooperative Agreement Number NNA 13AA93A. A.D.-G. was supported by NExSS. S.D. was supported by NASA Exobiology grant NNX15AM07G. J.K.-T. was supported by NASA Earth and Space Sciences Fellowship NNX15AR63H. A.J.R. was supported by an appointment to the NASA Postdoctoral Program at NASA Ames Research Center, administered by Universities Space Research Association under contract with NASA. Thanks to Ewan Cameron (University of Oxford) for making substantive suggestions about Bayesian statistics in an earlier draft, which greatly improved the final article. Thanks also to Duncan Forgan, Lee Grenfell, Sarah Rugheimer, and Eva Stüeken for extensive comments on the entire article, and Stephen Kane, Abel Mendez, and Vladimir Airapetian for additional comments on the NASA Nexus for Exoplanet System Science (NExSS) open community forum. We also thank NASA for their support for the NExSS Exoplanet Biosignatures Workshop in 2016 and Mary Voytek for her leadership of NExSS and feedback about the workshop and articles.

**Author Disclosure Statement**

No competing financial interests exist.

EXOPLANET BIOSIGNATURES ASSESSMENT											25

Address correspondence to:
*David C. Catling*
*Astrobiology Program*
*Department of Earth and Space Sciences*
*University of Washington*
*Box 351310*
*Seattle, WA 98103*

*E-mail:* dcatling@uw.edu

**Abbreviations Used**

HZ = habitable zone
IPCC = Intergovernmental Panel on Climate Change
IR = infrared
MODIS = Moderate Resolution Imaging Spectrometer
NDVI = normalized difference vegetation index
NIR = near-infrared
ppmv = parts per million by volume
SED = spectral energy distribution
SETI = search for extraterrestrial intelligence
TM = Thematic Mapper